\documentclass[galaxies,article,accept,pdftex,moreauthors]{Definitions/mdpi} 
\firstpage{1} 
\pubvolume{1}
\issuenum{1}
\articlenumber{0}
\pubyear{2022}
\copyrightyear{2022}
\datereceived{} 
\dateaccepted{} 
\datepublished{} 
\hreflink{https://doi.org/} 
\usepackage{float}
\usepackage{xcolor}
\usepackage{bm}
\DeclareRobustCommand{\ion}[2]{\textup{#1\,\textsc{\lowercase{#2}}}}

\Title{Unveiling the evolutionary state of three B supergiant stars: PU\,Gem, $\epsilon$\,CMa and $\eta$\,CMa}

\TitleCitation{\textbf{Unveiling the evolutionary state of three BSG stars}.}

\Author{Julieta P. S\'anchez Arias $^{1}$*\orcidA{}, P\'eter N\'emeth $^{1,2}$\orcidB{}, Elisson S. G. de Almeida$^{3}$\orcidE{}, Matias A. Ruiz Diaz  $^{4}$\orcidF{},  Michaela Kraus $^{1}$\orcidC{} and Maximiliano Haucke$^{5}$ }

\AuthorNames{Julieta P. S\'anchez Arias, P\'eter N\'emeth, Elisson S. G. de Almeida, Matias A. Ruiz Diaz,  Michaela Kraus and Maximiliano Haucke }

\AuthorCitation{S\'anchez Arias, J. P.; Nemeth, P.; de Almeida, E. S. G.; Ruiz Diaz, M. A.; Kraus, M. and Haucke, M.}

\address{%
$^{1}$ \quad Astronomical Institute, Czech Academy of Sciences, Fri\v{c}ova 298, 25165 Ond\v{r}ejov, Czech Republic\\
$^{2}$ \quad Astroserver.org, F\H{o} tér 1, 8533 Malomsok, Hungary\\
$^{3}$ \quad Instituto de Física y Astronomía, Universidad de Valparaíso. Av. Gran Bretaña 1111, Casilla 5030, Valparaíso, Chile.\\
$^{4}$ \quad Instituto de Astrof\'isica de La Plata. CONICET-UNLP, Argentina.  Paseo del Bosque s/n, La Plata, Argentina\\
  $^{5}$ \quad Universidad Nacional Arturo Jauretche, Buenos Aires, Argentina}
\corres{Correspondence:sanchez\makeatletter asu.cas.cz}

\abstract{
We aim to combine asteroseismology, spectroscopy, and evolutionary models to establish a comprehensive picture of the evolution of Galactic blue supergiant stars (BSG). To start such an investigation, we selected three BSG candidates for our analysis: HD\,42087 (PU\,Gem), HD\,52089 ($\epsilon$\,CMa) and HD\,58350 ($\eta$\,CMa). These stars show pulsations and were suspected to be in an evolutionary stage either preceding or succeding the red supergiant (RSG) stage.
For our analysis, we utilized the 2-min cadence TESS data to study the photometric variability and we obtained new spectroscopic observations at the CASLEO observatory. We used non-LTE radiative transfer models calculated with CMFGEN to derive their stellar and wind parameters. For the fitting procedure, we included CMFGEN models in the iterative spectral analysis pipeline {\sc XTgrid} to determine their CNO abundances. The spectral modeling was limited to changing only the effective temperature, surface gravity, CNO abundances, and mass-loss rates. Finally, we compared the derived metal abundances with prediction from Geneva stellar evolution models.
The frequency spectra of all three stars show stochastic oscillations and indications of one nonradial strange mode, $f_r=$ 0.09321 d$^{-1}$ in HD\,42087 and a rotational splitting centred in $f_2=$ 0.36366 d$^{-1}$  in HD\,52089. We conclude that the rather short sectoral observing windows of TESS prevent establishing a reliable mode identification of low frequencies connected to mass-loss variabilities. The spectral analysis confirmed gradual changes in the mass-loss rates and the derived CNO abundances comply with the values reported in the literature. We were able to achieve a quantitative match with stellar evolution models for the stellar masses and luminosities. However, the spectroscopic surface abundances turned out to be inconsistent with the theoretical predictions. The stars show N enrichment, typical for CNO cycle processed material, but the abundance ratios are not reflecting the associated levels of C and O depletion. We found HD\,42087 (PU\,Gem) to be most consistent with a pre-RSG evolutionary stage, and HD\,58350 ($\eta$\,CMa) is most likely in a post-RSG evolution. We were unable to model HD\,52089 ($\epsilon$\,CMa) with any evolved BSG tracks and conclude its relatively low mass and high luminosity are most consistent with a rejuvenated merger scenario.}

 
\begin{document}




\section{Introduction}

Massive stars are one of the most important objects in the universe due to their key role in the enrichment of interstellar medium with metals for future star generations and the evolution of the host galaxies \citep{1982ApJ...263..723A}. Nevertheless, understanding their evolution is challenging due to the significant changes they experience at different evolutionary stages, especially during the post-Main Sequence (MS). Therefore, any inaccuracy in the input parameters results in large uncertainties of the evolutionary models \citep{2013A&A...560A..16M, 2022MNRAS.512.5717A}. Fortunately, there are many sophisticated stellar evolutionary models calculated for massive stars \citep[e.g.][]{2022MNRAS.511.2814Y, 2010NewAR..54...32M}. However, they depend on internal parameters such as rotation, chemical mixing, and angular momentum transport, for which no decisive observational constraints are available. In addition, massive supergiant
stars undergo mass loss via line-driven winds and their rates are far from being firmly established, adding even more uncertainties to the evolution of these stars.


BSGs comprise extreme transition phases, in which the stars shed huge amounts of material into their environments. During the evolution of massive stars, those stars with masses between 20 and 40 $M_{\odot}$ evolve back to the blue supergiant state, after a Red Supergiant (RSG) state, either as Blue Supergiant (BSG) or in a follow-up Wolf-Rayet phase. On the other hand, massive stars with $< 20$ $M_{\odot}$ may experience "blue loops", where the star
changes from a cool star to a hotter one before
cooling again. Therefore, hot BSGs can be found at the pre-RSG stage, burning only H in a shell or at the post-RSG during the He-core burning. The exact reason why some massive stars experience  "blue loops" and others do not
still remains unknown, although it is known that
extra mixing processes within the layers surrounding their convective core along with mass loss events during their evolution play an essential role \citep{2019ApJ...886...27W, 2019NatAs...3..760B, 2012ARA&A..50..107L}.

In addition, BSGs can show an extremely rich spectrum of stellar oscillations. Since these oscillations depend on the internal structure of the star, their analysis paves the way to understanding phenomena such as the occurrence of “blue loops” or the discernment between the different evolutionary stages in which blue
supergiant stars can exist. The observed oscillation modes in B  supergiants are driven by different excitation mechanisms,
such as the classical $\kappa$ mechanism, stochastic wave generation caused by the presence
of convective layers in the outer regions of the star or in their envelope, tidal excitation causing the so-called Rossby modes and strange modes. Strange modes are nonlinear instabilities that require a luminosity over a mass ratio of 
$L_{\star}/M_{\star} > 10^4 L_{\odot} /M_{\odot}$ to be excited \citep{1994MNRAS.271...66G}. Their existence has been related to the variable mass loss these stars experience \citep{2015A&A...581A..75K, 2018A&A...614A..91H, 2010A&A...513L..11A}. 

Recent studies on stellar oscillations of individual BSGs \citep{2006ApJ...650.1111S, 2010A&A...513L..11A, 2015A&A...581A..75K} improved our understanding of the complex variability these stars show. Furthermore, considerable progress has been made in theoretical studies on the stellar oscillations of massive stars at different evolutionary stages focusing on specific mass ranges. For example, \cite{2015MNRAS.447.2378O} investigated stellar oscillations in stars at pre- and post-RSG stages for masses between 13 and 18 solar masses for different physical parameters such as metallicity and overshooting. In \cite{2013MNRAS.433.1246S}, a thorough study of stellar oscillations is presented for a broader range of masses, revealing that the pulsation properties in pre- and post-RSG evolutionary phases are fundamentally different: stars in the post-RSG stage pulsate in many more modes, including radial strange modes, than their less evolved counterparts, although these authors did not inspect the effect of different wind efficiency on the stability of the modes. 

Despite the efforts accumulated over years in the study of the stellar pulsations of these objects in different evolutionary stages, the correct identification of the evolutionary stage of BSGs additionally requires a detailed analysis of spectral observations due to the large uncertainties these stars have in astrophysics parameters such as the mass and the radii. 


Therefore, we started a comprehensive study of these objects with the aim of gaining insights into the evolutionary state of BSG stars. Combining information about the pulsation behaviour extracted from photometric lightcurves with newly determined stellar parameters and precise chemical abundances obtained from modeling of acquired spectroscopic data, we strive to find clear evidence for either a pre- or a post-RSG state of the objects. The current paper presents our methodology and first results on a small sample of objects.
 



The paper is organized as follows: in Sect. \ref{Peter} we summarize the main parameters for these stars found in the literature and Sect. \ref{observations} describes the spectroscopic and photometric observations employed in this work. In Sect. \ref{frequency_analysis}, we present the analysis of the light curves and the frequency spectra. Sect. \ref{peterandelisson} is devoted to the spectral analysis of the selected objects. We describe the numerical tools employed, including brand new capabilities implemented in {\sc XTgrid} \citep{2012MNRAS.427.2180N} for this work and the results obtained.  Finally, Sect. \ref{discussion} and \ref{conclusions} are devoted to the discussion and conclusions.

\section{Target selection and parameters values from literature}
\label{Peter}

\begin{figure}[t]
\centerline{\resizebox{0.80\textwidth}{!}{\includegraphics{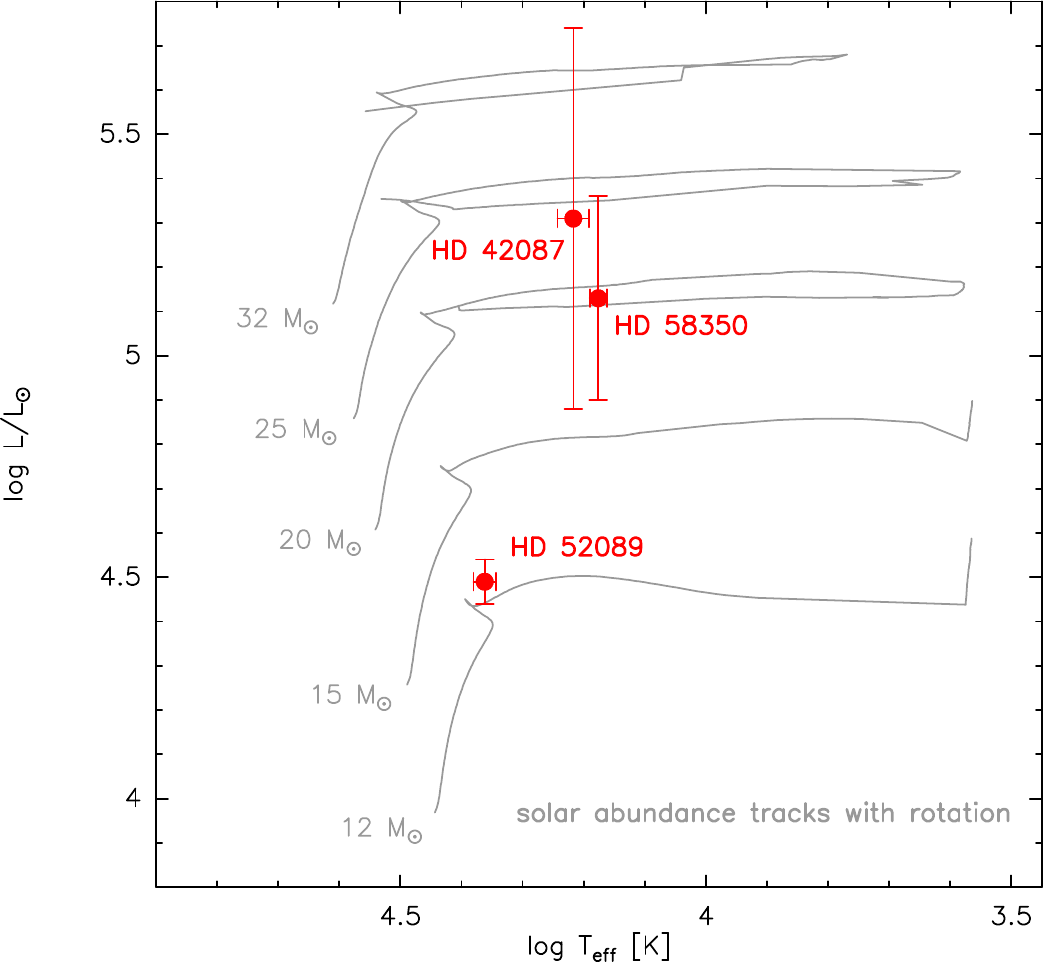}}}
\caption{Our three BSG stars in the HR diagram according to the values derived in \cite{2018A&A...614A..91H}. The evolutionary tracks are taken from \citep{2012A&A...537A.146E}}\label{HR}
\end{figure}

\citet{2018A&A...614A..91H} presented the most recent comprehensive analysis of the 
wind properties of 19 pulsating BSGs. We focus on three objects from their sample, for which we have obtained new spectroscopic observations. These are the stars HD\,42087 (PU Gem), HD\,52089 ($\epsilon$\,CMa) and HD\,58350 ($\eta$\,CMa). Table \ref{starstelpar} summarizes the values for these stars derived by \citet{2018A&A...614A..91H}. They serve as reference (or starting) values for our analysis. In Fig. \ref{HR} we show the position of the selected stars in an HR diagram for the parameters in Table \ref{starstelpar}. We notice that these hot BSG stars could be either at the immediate post-Main Sequence or at the post-Red Supergiant stage. 

The three selected objects have been studied extensively in the literature, and a compilation of literature values for their stellar and wind parameters can be found in \citet{2018A&A...614A..91H}. Here we restrict to a brief overview highlighting a variety of stellar and wind parameters.

\subsection{HD\,42087}
  \citet{searle08} derived the following parameters for this star using spectra from October 1990: $T_{\rm eff}$ = 18000 $\pm$ 1000 K, $\log g$ = 2.5, $\log(L_\star/L_{\odot})$ = 5.11 $\pm$ 0.24, $R_\star$ = 36.6 $R_{\odot}$, and $v \sin i$ = 71 km s$^{-1}$. They employed CMFGEN complemented with TLUSTY to derive $T_{\rm eff}$ and $\log(L_\star/L_{\odot})$, $\log g$, along with the CNO abundances being $\epsilon(C)$ = 7.76, $\epsilon(N)$ = 8.11, and $\epsilon(C)$ = 8.80. For the wind parameters, they obtained $\dot{M}$ = $5.0\times 10^{-7}$ $M_\odot$ yr\textsuperscript{-1}, $\beta$ = 1.2, and $v_\infty$ = 650 km s$^{-1}$.~\citet{morel04} showed that HD 42087 has a high H$\alpha$ variability with a spectral variability index (as defined by these authors) of $\sim$91\% in this line ~\citep[see Sect. 4 of][]{morel04}, evidencing a cyclic behaviour with P $\sim$25 d. We also mention that the values derived for this object in \citet{2018A&A...614A..91H} were obtained with TLUSTY using an optical spectrum covering only the H${\alpha}$ region. This spectra from 2006, showed a P Cygni feature with a weak emission and a strong absorption component.

\subsection{HD\,52089}
In \citet{2008A&A...481..453M} this star was studied as a slowly rotating B-type dwarf star. They derived $T_{\rm eff}$ = 23000 K and $\log{g}$ = 3.30 $\pm$ 0.15  using spectroscopic data from April 2005. By using the DETAIL/SURFACE code they determined the non-LTE abundances $\epsilon(C)$ = 8.09 $\pm$ 0.12, $\epsilon(N)$ = 7.93 $\pm$ 0.24, and $\epsilon(O)$ = 8.44 $\pm$ 0.18. \cite{2015A&A...574A..20F} derived an effective temperature and surface gravity of $T_{\rm eff}$ = 22 500 $\pm$ 300 K and $\log g$ = 3.40 $\pm$ 0.08. They also obtained updated values for the surface abundances by analyzing a FEROS spectra from 2011 and found $\epsilon(C)$ = 8.30 $\pm$ 0.07, $\epsilon(N)$ = 8.16 $\pm$ 0.07 and $\epsilon(O)$ = 8.70 $\pm$ 0.12.  Additionally, they estimated a 12.5 $M_{\odot}$ for this object.

\subsection{HD\,58350}

\citet{2007A&A...463.1093L} derived $T_{\rm eff}$ = 13500 K, $\log g$ = 1.75, $R_{\star}$ = 65 $R_{\odot}$, $\log (L_\star/L_{\odot})$ = 5.10, $v_\infty$ = 250 km s$^{-1}$, $\dot{M}$ = $1.4\times 10^{-7}$ $M_\odot$ yr\textsuperscript{-1} and $\beta$ = 2.5 for the stellar and wind parameters of this star. In \citet{searle08}, they derived $T_{\rm eff}$ = 15000 $\pm$ 500 K, $\log g$ = 2.13, $R_{\star}$ = 57.3 $\pm$ 2.64 $R_{\odot}$, and $\log (L_\star/L_{\odot})=$ 5.18 $\pm$ 0.17. For the CNO abundances they obtained $\epsilon(C)$ = 7.78, $\epsilon(N)$ = 8.29, and $\epsilon(O)$ = 8.75.

\begin{table}[H] 

\caption{Stellar and wind parameters derived by \citet{2018A&A...614A..91H} for our star sample.} \label{starstelpar}
\newcolumntype{C}{>{\centering\arraybackslash}X}
\begin{tabularx}{\textwidth}{lCCC}
\toprule
\textbf{Parameter}                                      & \textbf{HD\,42087}& \textbf{HD\,52089}& \textbf{HD\,58350} \\  
\textbf{$T_{\rm eff}$} [K]            & 16500 $\pm$ 1000    & 23000 $\pm$ 1000    & 15500 $\pm$ 700      \\
\textbf{$\log g$} [cgs]	              & 2.45 $\pm$ 0.10     & 3.00 $\pm$ 0.10     & 2.00 $\pm$ 0.10      \\
\textbf{$\log{L_\star}$} [$L_{\odot}$]& 5.31 $\pm$ 0.43     & 4.49 $\pm$ 0.05     & 5.18 $\pm$ 0.32      \\
\textbf{$R_\star$} [$R_{\odot}$]      & 55                & 11                & 54                 \\
$v\sin i$ [km s\textsuperscript{-1}]  & 80                & 10                & 40                 \\
\midrule
\textbf{$\dot{M}$} [$M_\odot$ yr\textsuperscript{-1}]& (5.7 $\pm$ 0.5)$\times10^{-7}$&(2.0 $\pm$ 0.6)$\times10^{-8}$&(1.4 $\pm$ 0.2)$\times10^{-7}$     \\
$v_\infty$ [km s\textsuperscript{-1}] & 700 $\pm$ 70        & 900 $\pm$ 270       & 200 $\pm$ 30         \\
$\beta$                               & 2.0               & 1.0               & 3.0                \\

\bottomrule

\end{tabularx}
\end{table}

\section{Observations}
\label{observations}

With the aim to shed light on the evolutionary state of these objects, we analyze new spectroscopic observations and combine these results with information extracted from their photometric light curves. 
\subsection{Spectra}
 The spectra we employed in this work to derive new parameters were taken on 2020 January 23 and 24 for HD\,42087 and HD\,58350, respectively, and 2015 February 14 for HD\,52089. All of them cover the wavelength range from 4275 \AA\ up to 6800 \AA\ with a signal-to-noise ratio S/N of 140, 140, and 130 for HD\,42087, HD\,52089, and HD\,58350, respectively.

We utilized the REOSC spectrograph attached to the {\it{Jorge Sahade}} 2.15 m telescope at the Complejo Astron\'omico El Leoncito (CASLEO), San Juan, Argentina. 
The resolving power at 4500 \AA\ and 6500 \AA\ is $R \sim 12600$ and $R \sim 13900$, respectively. 
The spectra were reduced and normalized following standard procedures using IRAF\footnote{IRAF is distributed by the National Optical Astronomy Observatory, which is operated by the Association of Universities for Research in Astronomy (AURA) under a cooperative agreement with the National Science Foundation.} routines.

\begin{figure}[H]
\begin{adjustwidth}{-\extralength}{-5cm}
\centering
\includegraphics[width=0.7\textwidth]{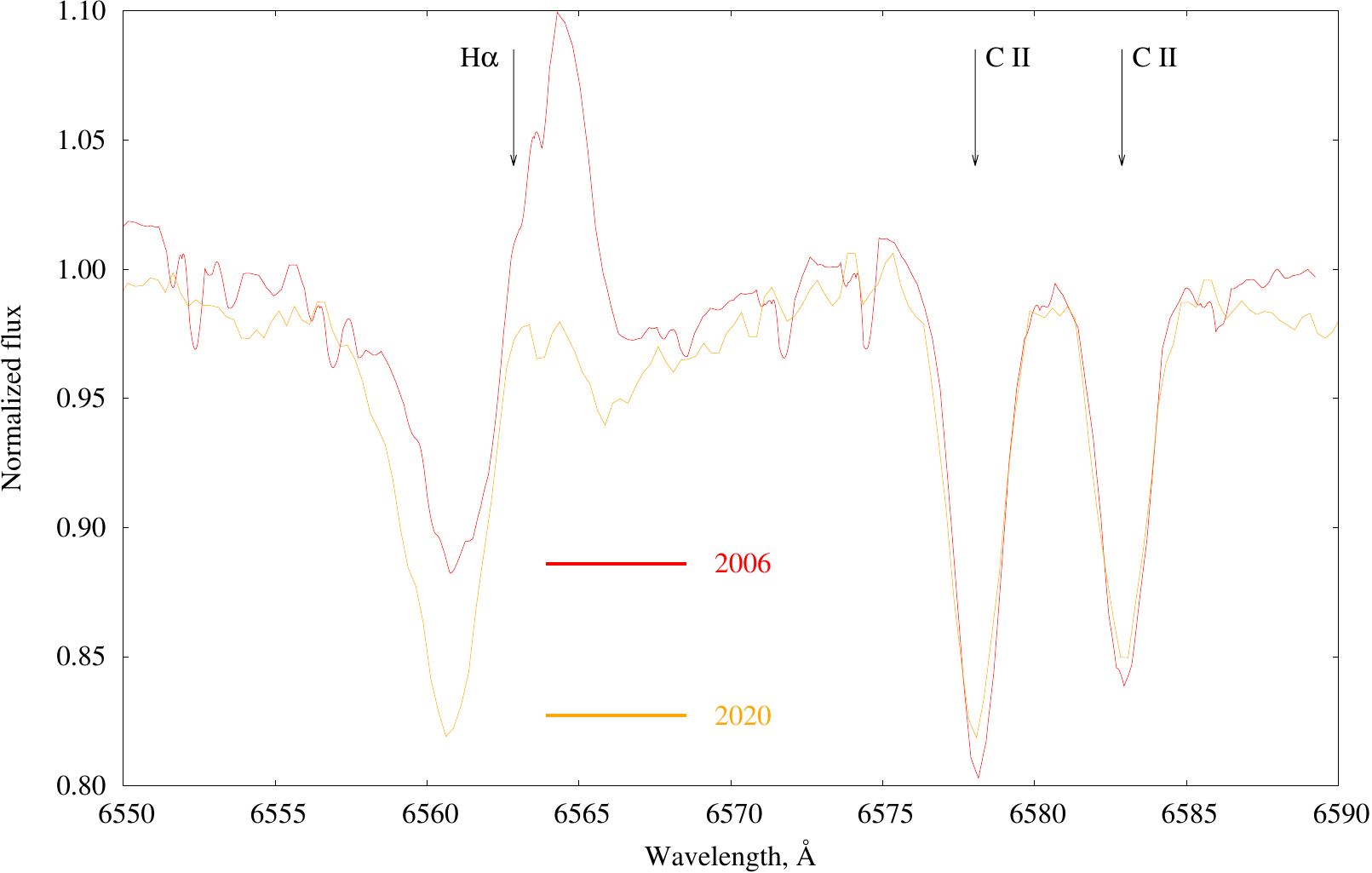}
\end{adjustwidth}
\caption{Evolution of the H$\alpha$ line profile of HD\,42087 between 2006 and 2020. 
The emission component  weakened by 2020. 
\label{fig_HD42_alpha}}
\end{figure}

\begin{figure}[H]
\begin{adjustwidth}{-\extralength}{-5cm}
\centering
\includegraphics[width=0.7\textwidth]{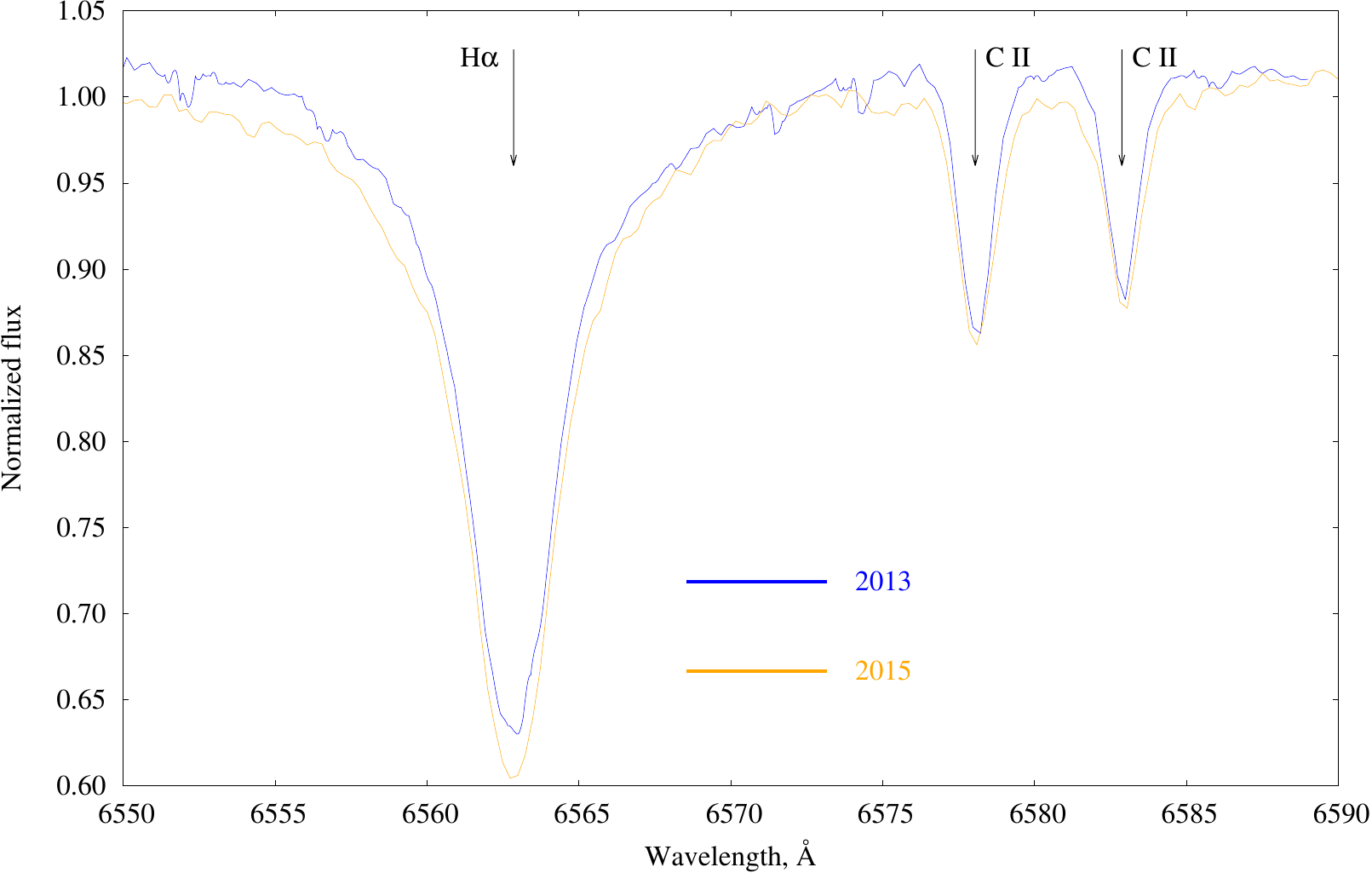}
\end{adjustwidth}
\caption{Evolution of the H$\alpha$ line profile of HD\,52089 between 2013 and 2015. 
\label{fig_HD52_alpha}}
\end{figure}

\begin{figure}[H]
\begin{adjustwidth}{-\extralength}{-5cm}
\centering
\includegraphics[width=0.7\textwidth]{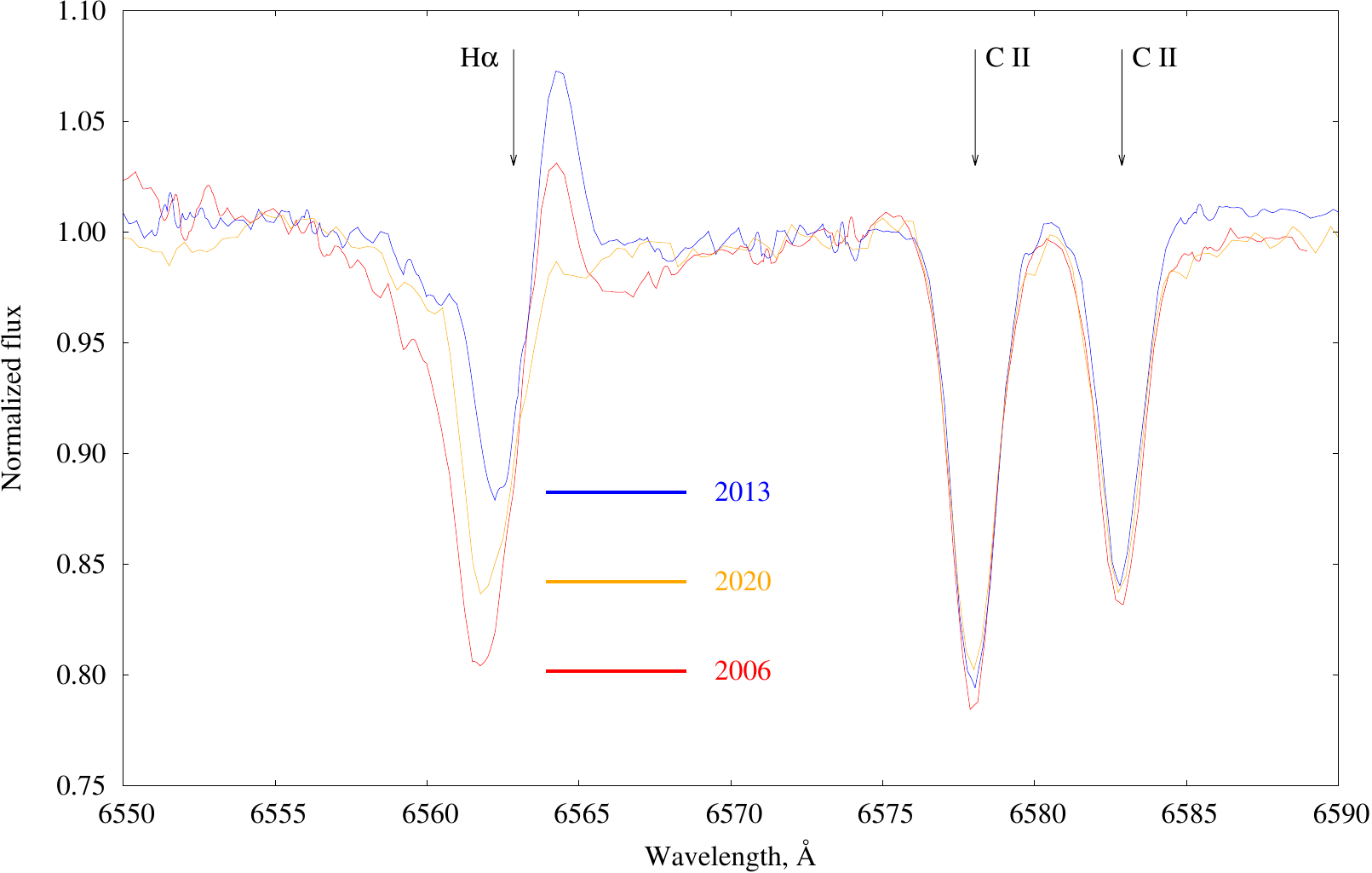}
\end{adjustwidth}
\caption{Evolution of the H$\alpha$ line profile of HD\,58350 between 2006 and 2020. 
The emission component has strengthened between 2006 and 2013 and got noticeably weaker by 2020. 
\label{fig_HD58_alpha}}
\end{figure}

In order to show the H$\alpha$  variability in our targets, we have collected previous observations in CASLEO, depicted in Fig. \ref{fig_HD42_alpha}, \ref{fig_HD52_alpha}, and \ref{fig_HD58_alpha}; from 2006 January 15 for HD 42087; 2013 February 5 for HD 52089; and 2006 January 15 and 2013 February 5 for HD 58350.

\label{Juliymati}

\subsection{Photometric light curves}
We complemented the optical spectra with space photometry collected with the Transiting Exoplanet Survey Satellite (TESS) mission \citep{2014SPIE.9143E..20R,2015JATIS...1a4003R}. We retrieved high cadence (120s) PDCSAP light curve for each object using Astroquery \citep{2019AJ....157...98G} and cleaned the light curves using the Lightkurve package \citep{2018ascl.soft12013L}. For this, we first selected only the data points with a quality flag equal to 0, meaning they are not cosmic rays or data from bad pixels. Next, we carried out sigma clipping to remove outliers, using a value of 6$\sigma$ for all light curves, following a similar procedure as in \citet{2022A&A...662A..82G}. The resulting light curves were transformed to display the variation of the magnitude ($\Delta m$) \textbf{from} the mean magnitude:

\begin{equation}
\Delta m = -2.5 \mathrm{log}(\mathrm{pdcsap\_flux}) + 2.5 \mathrm{log}(\overline{\mathrm{pdcsap\_flux}}) ; 
\end{equation}

\noindent where pdcsap\_flux is the individual flux at each exposure and $\overline{\mathrm{pdcsap\_flux}}$ is the mean flux of the entire light curve. 
In this way, we obtain a normalized light curve, whose amplitude can be expressed in units of $\mathrm{mag}$. The time axis is in units of Barycentric Julian Date, which is the Julian Date corrected for differences in the earth's position with respect to the barycenter of the solar system. 

\section{Frequency analysis}
\label{frequency_analysis}
We employed the Fourier Transform with Period04 \citep{2005CoAst.146...53L}. For each star and sector, the frequencies were searched in the interval [0;50] $d^{-1}$, widely covering their frequency content. No frequency beyond 2 $d^{-1}$ was found for any star. The amplitude and phase were calculated using a least square sine fit for each detected frequency. After obtaining the first frequency, the analysis was performed on the residuals. The Fourier analysis was stopped after obtaining 15 frequencies. Once we derived the frequencies following this procedure, we dismissed those frequencies below 0.1 $d^{-1}$ since TESS data of a single sector do not allow us to derive periods higher than $\sim 10$ d. Additionally, we discard those frequencies with a separation of less than $2/T$, where $1/T$ is the Rayleigh resolution and $T$ is the time span of the observations. The S/N ratio was computed for the derived frequencies along with the uncoupled uncertainties in the frequencies and amplitudes using a Monte Carlo simulation.

When available, we also analyzed the frequency content of the combined, consecutive sectors since the longer time baseline would facilitate the 
detection of longer periods from radial modes possibly connected to strange modes. We considered the values recommended in~\citet{2021AcA....71..113B} for the S/N when dealing with individual TESS sectors and combined sectors, which resulted in 5.037 for our individual TESS sectors, and of 5.124 and 5.194 for the combined sectors of HD\,52089 (2 sectors combined) and HD\,42087 (3 sectors combined), respectively. Nevertheless, we took into account frequencies with lower S/N whenever they turned out to be interesting for the analysis (see below).

Finally, when comparing frequencies from different sectors, we adopted the separation criterion for the combined sectors.

Next, we provide the details of the light curves and the frequencies extracted for each star.

\subsection{HD\,42087}
\begin{figure}[H]
\begin{adjustwidth}{-\extralength}{-5cm}
\centering
\includegraphics[width=16.5cm]{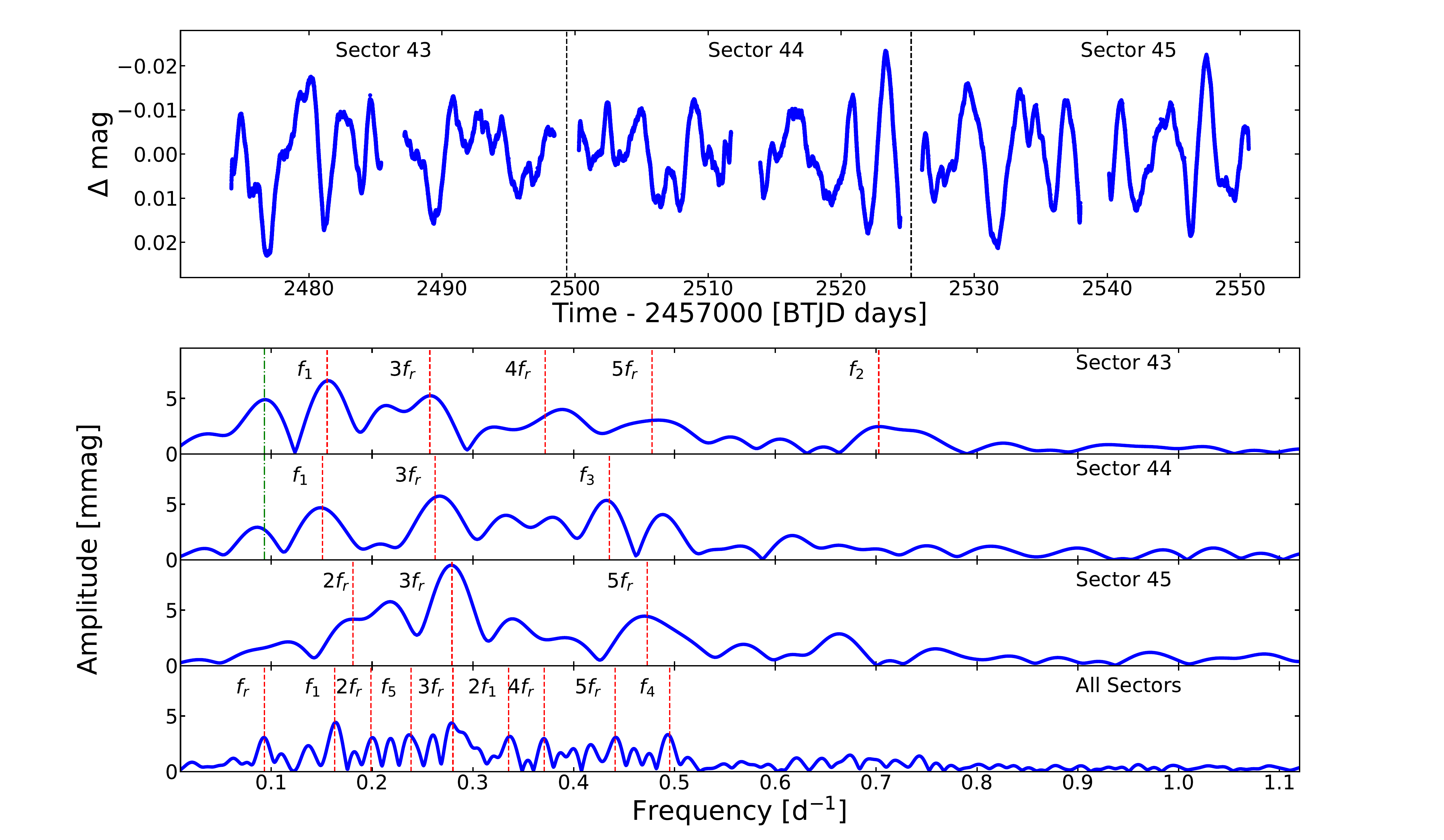}
\end{adjustwidth}
\caption{{\it{Top panel:}} light curve of HD\,42087 corresponding to Sectors 43, 44, and 45. {\it{Lower panels:}} Amplitude spectra for each sector and all sectors combined. \label{fig7}}
\end{figure}

This star was observed in 3 consecutive sectors: Sector 43, in the period 2021 September 16 to October 10; Sector 44, during 2021 October 10 to November 6; and Sector 45, during 2021 November 6 to December 2, covering an observation time span of 24.287 days, 24.156 days and 24.551 days, respectively. The light curve and the amplitude spectra for the individual sectors and all sectors combined are displayed in Fig. \ref{fig7}.

The frequencies derived for sectors 43, 44, and 45 are listed in Table \ref{tab4} and the threshold considered for the frequency separation is $0.0823$ $d^{-1}$, $0.0828$ $d^{-1}$ and $0.0815$ $d^{-1}$, respectively. For the analysis of all sectors combined, the time span is $T$ = 72.994 $d$, therefore, the separation criterion we employed is $2/T$ = 0.0274 $d^{-1}$. We adopted this value as a conservative threshold when comparing frequencies from different sectors.

The pulsational content of HD\,42087 lies at low frequencies below 0.8 $d^{-1}$ and the amplitudes are between 2.5 and 9.5 mmag, approximately. When analyzing the combined sectors, we also searched for frequencies down to 0.025 $d^{-1}$ due to the extended length of the observations. This allowed us to find the frequency $f_r$ and its five harmonics. Some of these harmonics appear in the individual sectors. Due to the short time span for individual sectors, frequencies below $\sim$0.1 $d^{-1}$ are not reliable; nevertheless, $f_r$ seems to appear for sectors 43 and 44 (indicated with green lines in Fig. \ref{fig7}). Besides $f_r$ and its harmonics, five frequencies appear randomly over the four sets of observations, which are likely related to stochastic oscillations (see Sect. \ref{discussion}).


\begin{table}[H] 
\caption{\textbf{List of frequencies, their amplitudes, and S/N ratios, found for  HD\,42087.}\label{tab4}}

\newcolumntype{C}{>{\centering\arraybackslash}X}
\begin{tabularx}{\textwidth}{CCCCCCC}
\toprule
\textbf{Sector} & \textbf{Frequency}& \textbf{3$\sigma_f$}	& \textbf{Amplitude}	& \textbf{3$\sigma_A$}& \textbf{S/N} & \textbf{Id} \\
& \textbf{[$d^{-1}$]} & \textbf{[$d^{-1}$]} &\textbf{[mmag]} & \textbf{[mmag]} & & \\
 \midrule
    & 0.15543 & 0.00044 & 7.7615 & 0.15 &10.65 & $f_1$\\
 43 & 0.25729 & 0.00063 & 5.7467 & 0.14 & 8.41 & $3*f_r$\\
    & 0.47759 & 0.00064 & 4.7663 &  0.13 & 7.79 & $5*f_r$\\
    &  0.37169 & 0.00090 & 3.5347 & 0.13 & 5.45 & $4*f_r$\\
    & 0.70257 & 0.00037 &2.9734 & 0.14 &5.41 & $f_2$\\

 

 \midrule
   & 0.26268  & 0.00068& 5.5317 & 0.16& 7.68 & $3*f_r$\\
44    & 0.15072 & 0.00082& 5.2203 & 0.17 &6.75 & $f_1$\\
 & 0.43536 & 0.00090& 4.9832 & 0.19 &7.45  & $f_3$\\
  

 \midrule
 & 0.27928 & 0.00009 & 9.2074 & 0.10& 10.65 & $ 3*f_r$\\
 45  & 0.18110   &  0.00059      &5.0188   & 0.13   & 5.57 & $2*f_r$\\
   & 0.47295  &  0.00159     & 4.0068  &   0.21 &5.07 & $5*f_r$\\
\midrule 

          & 0.28022 & 0.00019 & 4.4715 & 0.12 & 8.48 & $3*f_r$\\
 All & 0.16305 & 0.00021 & 3.9728 &0.11& 7.05 & $f_1$\\
    & 0.49515 & 0.00022  & 3.4982 &0.11 & 7.44 & $ f_4$\\ 
        & 0.23874 &  0.00027 & 3.2680 &  0.11 & 6.04  & $f_5$ \\
           &0.44102  &0.00023  &3.2321 &0.10 & 6.67 & $5*f_r$\\
   &0.33531  &    0.00030  & 2.9450    &      0.11  & 5.74   &  $2*f_1$  \\
     &0.19899  &   0.00029 & 2.8086  &    0.10    &   5.07 &   $2*f_r$    \\
     
     &0.37073 &  0.00033 & 2.6635  &  0.10  &  5.29 &  $4*f_r$ \\  
    
    & 0.09321 &  0.00032   & 2.4597  &     0.12   &  4.20  & $f_r$ \\

     
     
\bottomrule

\end{tabularx}
\end{table}

\subsection{HD\,52089}
\begin{figure}[H]
\begin{adjustwidth}{-\extralength}{-5cm}
\centering
\includegraphics[width=16.5cm]{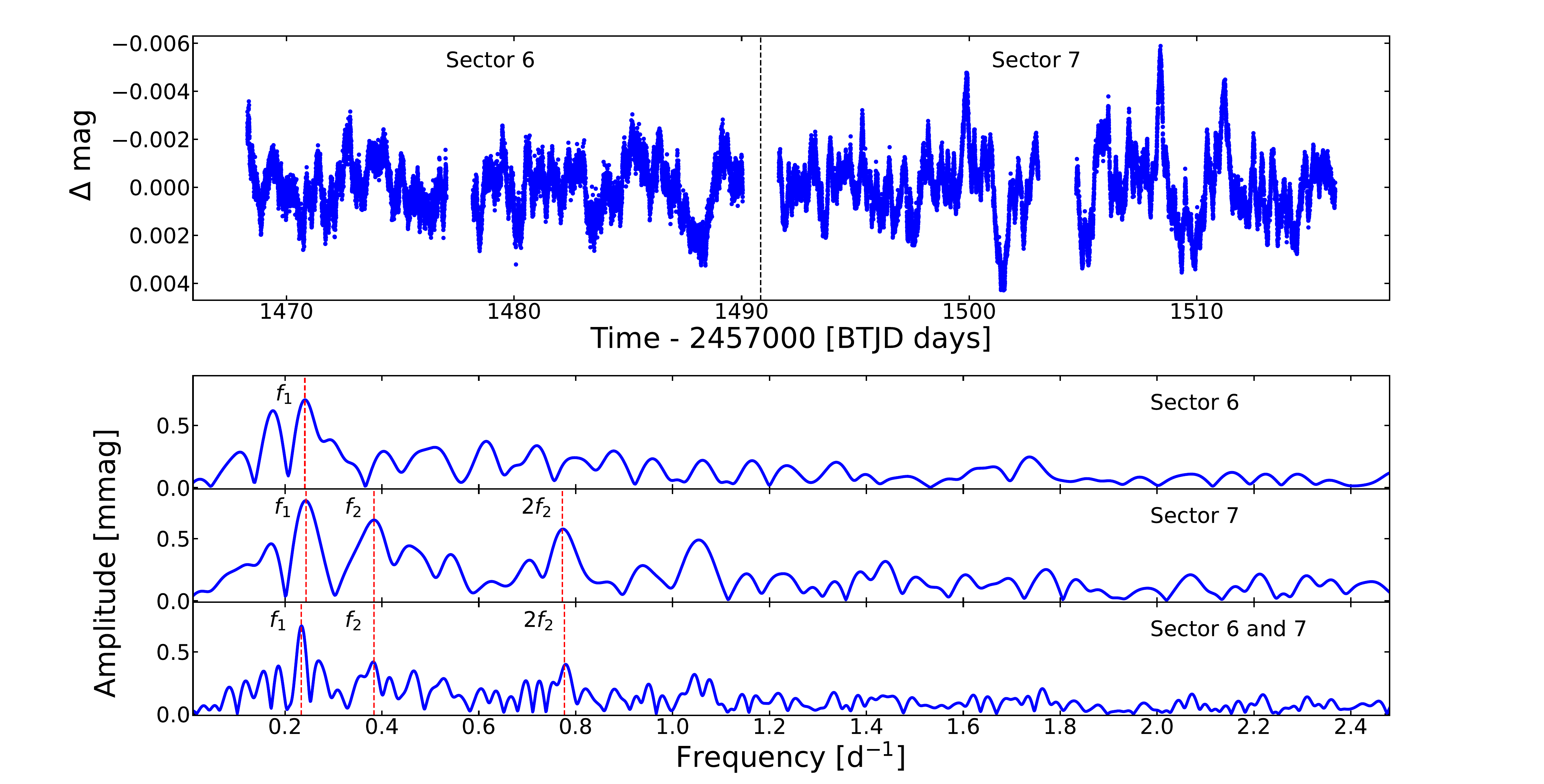}
\end{adjustwidth}
\caption{{\it{Top panel:}} light curve of HD\,52089 corresponding to Sectors 6 and 7. {\it{Lower panels:} } Amplitude spectra for each sector and all sectors combined. \label{fig8}}
\end{figure}

\begin{figure}[H]
\begin{adjustwidth}{-\extralength}{-5cm}
\centering
\includegraphics[width=16.5cm]{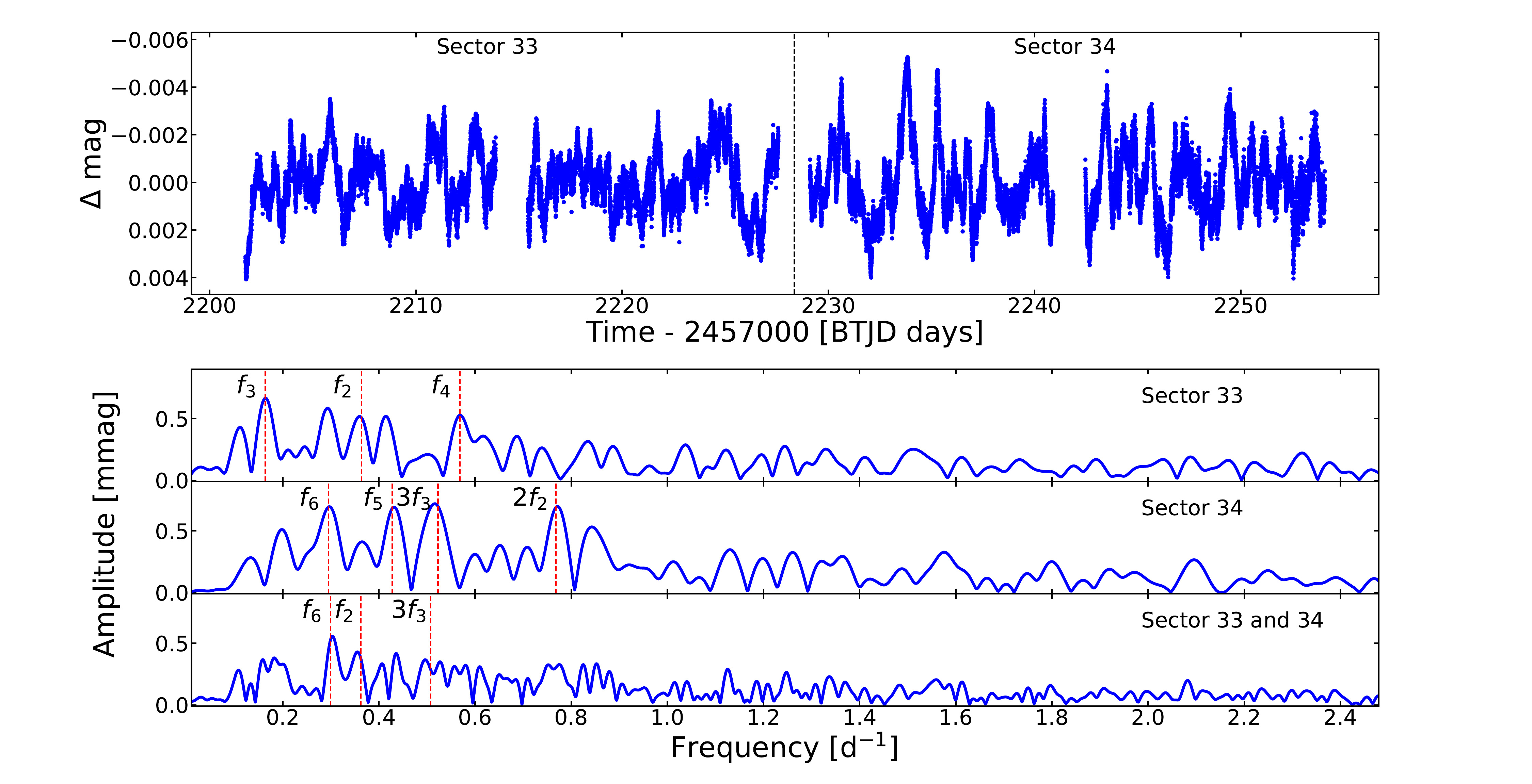}
\end{adjustwidth}
\caption{{\it{Top panel:}} light curve of HD\,52089 corresponding to Sectors 33 and 34. {\it{Lower panels:} } Amplitude spectra for each sector and all sectors combined. \label{fig9}}
\end{figure}

 This star was observed in 4 TESS sectors: Sector 6, in the period 2018 December 15 to 2019 January 6; Sector 7, in the period 2019 January 8 to February 1; Sector 33 from 2020 December 18 to 2021 January 13; and Sector 34 in the period 2021 January 13 to February 8. The total time spans for Sectors 6, 7, 33, and 34 are 21.771 d, 24.454 d, 25.839 d, and 24.962 d, respectively. Fig. \ref{fig8} displays the light curves and amplitude spectra for Sectors 6 and 7, and Fig.\ref{fig9} displays the same information for Sectors 33 and 34. Due to the large time gap between sectors 7 and 33, we decided to evaluate the sectors in pairs, i.e. together with the analysis of sectors 6 and 7, we studied the frequencies of these sectors combined, and the same with sectors 33 and 34. For the combined sectors 6 and 7, the time span is 46.225 d, and for sectors 33 and 34 together, the time span is 50.801 d.

The derived frequencies, their amplitudes and identifications are listed in Table \ref{tab10}. The separation criteria we employed are  0.0918 $d^{-1}$, 0.0817 $d^{-1}$, 0.0774 $d^{-1}$, and 0.0801 $d^{-1}$ for sectors 6, 7, 33, and 34, respectively. For sectors 6 and 7 combined and 33 and 34 combined the separation criteria are 0.0432 $d^{-1}$ and 0.0393 $d^{-1}$, respectively and the lowest detectable frequency for the sectors combined is $\sim$0.04 $d^{-1}$. We adopted conservative separation criteria with 0.0393 $d^{-1}$ to compare frequencies from different sectors. We find only one significant frequency, $f_1$,  in Sector 6, which is also in Sector 7. In Sector 7, we found another frequency, $f_2$, which stays present also in Sector 33, and its harmonic, which appears in Sector 34. We searched for lower frequencies in both combined light curves, but we did not find any. After the gap of $\sim$2 years in the observations, new frequencies appear randomly in the new Sectors, which indicates they are not related to stellar oscillations due to their short lifetime ($f_3$, $f_5$, $f_6$). We note that $f_2-f_3 \sim f_4-f_2$ in Sector 33, indicating the presence of a triplet centred in $f_2$ with a possible rotation frequency of 0.20 $d^{-1}$. Nevertheless, the observations' short time span makes precise classification difficult. We did not find any significant frequency at lower ranges in the combined sectors. 



\begin{table}[H] 
\caption{Sector, Frequency, Amplitude, and Identification for HD\,52089.\label{tab10}}

\newcolumntype{C}{>{\centering\arraybackslash}X}
\begin{tabularx}{\textwidth}{CCCCCCC}
\toprule
\textbf{Sector} & \textbf{Frequency}& \textbf{3$\sigma_f$}	& \textbf{Amplitude}	& \textbf{3$\sigma_A$}& \textbf{S/N} & \textbf{Id} \\
& \textbf{[$d^{-1}$]} & \textbf{[$d^{-1}$]} &\textbf{[mmag]} & \textbf{[mmag]} & & \\

\midrule
 6 & 0.24083 & 0.00115& 0.7830& 0.05 & 6.15 & $f_1$\\


\midrule
  & 0.24316 & 0.01540& 0.8982 & 0.47&  6.14 & $f_1$\\
7 & 0.38381 & 0.00192& 0.7286  & 0.05&5.15 & $f_2$\\
  & 0.77252 & 0.00051& 0.5888 & 0.02& 4.37 & $ 2*f_2$\\
\midrule
         & 0.23321 & 0.00072& 0.7736 & 0.07 & 7.28 & $f_{1}$\\
 6 \& 7 & 0.38342 & 0.02775& 0.4930 & 0.17&4.79 & $f_{2}$\\
        & 0.77679 & 0.00073& 0.4769 & 0.04& 4.83 & $2*f_2$\\
\midrule
   & 0.16358 & 0.00070& 0.7716 & 0.02&6.868 & $f_3$\\
 33  & 0.36366 & 0.00089 & 0.5784 & 0.02 &5.42 & $ f_2$\\
   & 0.56843 & 0.00089& 0.5083 & 0.02& 4.72 & $f_4$\\

\midrule
    & 0.42781 & 0.00107& 0.7556 & 0.03& 4.98 & $f_{5}$\\
34   & 0.52280 & 0.00128& 0.7113 & 0.03 & 4.74 & $3*f_{3}$\\ 
   & 0.76818 & 0.00127& 0.7038 & 0.03& 4.75 & $ 2*f_2$\\
 & 0.29483 & 0.00147& 0.6389 & 0.03& 4.35 & $f_6$\\
\midrule
         & 0.29937 &0.00167& 0.6627 &0.12& 5.01 & $f_6$\\
33 \& 34 & 0.36220 & 0.00331& 0.5806 &0.34& 4.39 & $f_2$\\
     & 0.50772   &    0.00158  & 0.4714 &  0.09   & 4.09 &  $3*f_3$ \\
\bottomrule
\end{tabularx}
\end{table}

\subsection{HD\,58350}\label{HD58350LCs}

\begin{figure}[H]
\begin{adjustwidth}{-\extralength}{-5cm}
\centering
\includegraphics[width=16.5cm]{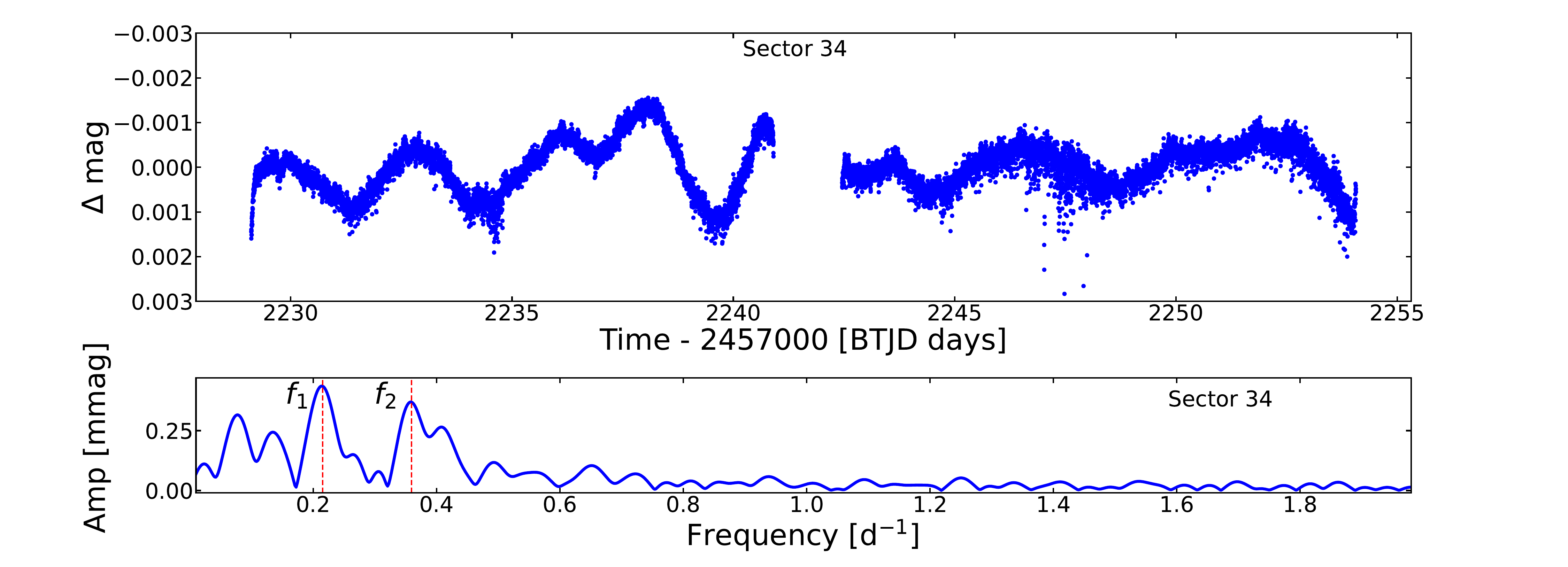}
\end{adjustwidth}
\caption{{\it{Top panel:}} light curve of HD\,58350 acquired during Sector 34. {\it{Lower panel:}} Amplitude \textbf{spectrum} for the same sector. \label{fig5}}
\end{figure}

This star has been observed in Sector 34 in the period 2021 January 14 to February 8, covering a total timespan of $T$ = 24.96 days. The light curve for sector 34 is displayed in the top panel of Fig. \ref{fig5} and  we show the amplitude spectra in the lower panel.

The resulting list of frequencies for HD\,58350 is shown in Table. \ref{tab10}. The separation criterion is $2/T$ = 0.0801 $d^{-1}$. The pulsational content for this BSG lies below 0.5 $d^{-1}$. We found 2 independent frequencies with similar amplitudes. 

\begin{table}[H] 
\caption{Sector, Frequency, Amplitude, and Identification for HD\,58350.\label{tab10}}

\newcolumntype{C}{>{\centering\arraybackslash}X}
\begin{tabularx}{\textwidth}{CCCCCCC}
\toprule
\textbf{Sector} & \textbf{Frequency}& \textbf{3$\sigma_f$}	& \textbf{Amplitude}	& \textbf{3$\sigma_A$}& \textbf{S/N} & \textbf{Id} \\
& \textbf{[$d^{-1}$]} & \textbf{[$d^{-1}$]} &\textbf{[mmag]} & \textbf{[mmag]} & & \\

\midrule
  34 &0.21533 & 0.00069& 0.4392 & 0.01& 8.54 & $f_1$\\
  &0.35927 & 0.00088& 0.3339 & 0.01& 7.07 & $f_2$\\
\bottomrule
\end{tabularx}
\end{table}

\section{Modeling tools for spectral analysis}
\label{peterandelisson}
\subsection{The code CMFGEN}
\label{elisson_code_cmfgen}

CMFGEN~\citep{hillier98} is one of the state-of-the-art 1-D non-LTE radiative transfer codes for studying the physical and chemical properties of massive hot stars. It has been successfully used for reproducing different types of observables of massive stars, such as in the case of OB supergiants: photometry, spectroscopy, and interferometry~\citep[e.g.,][]{searle08, bouret12, deAlmeida22}.\par

CMFGEN solves the radiative transfer and statistical equilibrium equations for the photosphere and wind under the assumption of radiative equilibrium and considers a spherically symmetric stationary wind. For the photospheric region, CMFGEN solves the radiative transfer and hydrostatic equations (describing the state of radiation and gas in the photosphere) in a self-consistent way. Further details on this procedure can be found in Sect.~3.1 of \citet{bouret13}.\par 

The wind velocity, $v(r)$, and density, $\rho(r)$, profiles are related to each other as follows, based on the assumption of a stationary spherical symmetric wind:

\begin{equation}
\dot{M} = 4\pi r^{2}\rho(r)v(r)f(r),
\label{eq:mass_continuity}
\end{equation}
where $r$ is the distance from the center of the star and $\dot{M}$ is the wind mass-loss rate, which is constant at any location of the wind. Here, the volume filling factor $f(r)$ parameterizes the inclusion of micro-clumping~\citep[e.g., see][]{hillier01}, that is, inhomogeneities that are found in the winds of massive stars~\citep[e.g.,][]{eversberg98, bouret05}. In CMFGEN, this parameterization is performed as follows:

\begin{equation}
f(r) = f_{\infty} + (1 - f_{\infty})\mathrm{e}^{-\frac{v(r)}{v_{\mathrm{initial}}}},
\label{eq:cmfgen_clumping}
\end{equation}
where $f_{\infty}$ is the filling factor value at $r\to\infty$ and $v_{\mathrm{initial}}$ is the onset velocity of clumping in the wind. Since our initial guesses, in the fitting procedure (as described in Sect.~\ref{peter_xtgrid}), for the stellar and wind parameters are based on the results from \citet{2018A&A...614A..91H}, we set $f_{\infty}$ = 1.0 in our default models. This means that our default CMFGEN models are calculated considering a homogeneous wind.\par

Furthermore, the wind's velocity profile is parameterized in CMFGEN by the so-called $\beta$-law approximation as follows:

\begin{equation}
v(r) = v_\infty\left(1 - \frac{R_\star}{r}\right)^{\beta},
\label{eq:beta_law}
\end{equation}
where $v_\infty$ is the wind terminal velocity and $R_\star$ is the stellar radius ($r$ higher than $R_\star$). For OB supergiants, values of $\beta$ are usually found to be as high as $\sim$2.0-3.0~\citep[e.g.,][]{martins15}. In short, Eqs.~\ref{eq:mass_continuity} and \ref{eq:beta_law} sets the relation between the most important fundamental wind physical parameters: the wind mass-loss rate and the terminal velocity.\par

\begin{table}[!ht]
\caption{\label{atomic_species} Summary on the atoms and ionization states of our default CMFGEN models: number of energy levels, super-levels, and bound-bound transitions for each atomic specie.
}
\centering
\renewcommand{\arraystretch}{1.1}
\begin{tabular}{lcccc}
\toprule
\toprule
Ion & Full-levels & Super-levels & b-b transitions \\
\midrule
\ion{H}{I} & 30 & 30  & 435\\

\ion{He}{I} & 69 & 69 & 905 \\
 \ion{He}{II} & 30 & 30 & 435 \\

\ion{C}{II} & 322 & 92 & 7742 \\
\ion{C}{III}  & 243 & 99 & 5528 \\
\ion{C}{IV}  & 64 & 64 & 1446 \\

\ion{N}{II} & 105 & 59 & 898 \\
\ion{N}{III} & 287 & 57 & 6223 \\
\ion{N}{IV} & 70 & 44 & 440 \\
\ion{N}{V} & 49 & 41 & 519 \\

\ion{O}{II} & 274 & 155 & 5880 \\
\ion{O}{III} & 104 & 36 & 761 \\
\ion{O}{IV} & 64 & 30 & 359 \\
\ion{O}{V} & 56 & 32 & 314 \\

\ion{Ne}{II} & 48 & 14 & 328 \\
\ion{Ne}{III} & 71 & 23 & 460 \\
\ion{Ne}{IV} & 52 & 17 & 315 \\

\ion{Mg}{II} & 44 & 36 & 348 \\

\ion{Si}{II} & 53 & 27 & 278 \\
\ion{Si}{III} & 90 & 51 & 640\\
\ion{Si}{IV} & 66 & 66 & 1090 \\

\ion{S}{III} & 78 & 39 & 520 \\
\ion{S}{IV} & 108 & 40 & 958 \\
\ion{S}{V} & 144 & 37 & 1673 \\

\ion{Fe}{II} & 295 & 24 & 2135 \\
\ion{Fe}{III} & 607 & 65 & 6670 \\
\ion{Fe}{IV} & 1000 & 100 & 37899 \\
\ion{Fe}{V} &1000 & 139 & 37737 \\
\ion{Fe}{VI} & 1000 & 59 &36431 \\

\ion{Ni}{II} &158 & 27 & 1668 \\
\ion{Ni}{III} &150 & 24 & 1345 \\
\ion{Ni}{IV} & 200 & 36 & 2337 \\
\ion{Ni}{V} & 183 & 46 & 1524 \\
\ion{Ni}{VI} & 182 & 40 & 1895 \\

\bottomrule
\end{tabular}
\end{table}

Despite its high computational cost, when compared to other non-LTE codes, such as FASTWIND, one of the biggest advantages of CMFGEN relies on allowing to set a complex chemical composition, with the inclusion of energy levels of different ions from hydrogen up to nickel. For instance,~\citet{2018A&A...614A..91H} studied our star sample of three BSGs employing radiative transfer models calculated with FASTWIND considering hydrogen, helium, and silicon with an approximation for the treatment of the line transfer of iron-group elements. In fact, the main difference between these two codes relies on an ``exact'' treatment of line-blanketing that is performed by CMFGEN~\citep[e.g., see][and references therein]{sander17}. Table \ref{atomic_species} summarizes the atomic species, the number of energy levels, and bound-bound transitions that are included as default in our models. Our models are calculated considering a robust atomic model for studying massive hot stars, including species of hydrogen, helium, carbon, nitrogen, oxygen, neon, magnesium, silicon, sulfur, iron, and nickel.\par

Finally, due to the complexity of the code and its computational cost, a very common approach when using CMFGEN is to vary its parameters manually in order to find an acceptable ``by eye'' fit to the observations~\citep[e.g.,][]{martins05, marcolino09, deAlmeida19, rivet20, deAlmeida22}. However, in this paper, we implement an automatic fitting procedure with CMFGEN to find the best-fit models for the observed spectrum of each star, as described below.

\subsection{Spectral analysis with {\sc XTgrid}}
\label{peter_xtgrid}

{\sc XTgrid} \citep{2012MNRAS.427.2180N} is a steepest-descent iterative $\chi^2$ minimizing fit procedure to model hot star spectra. 
The procedure was developed for the model atmosphere code {\sc Tlusty} \citep{HubenyLanz1995, LanzHubeny2007, HubenyLanz2017} and was previously applied to ultraviolet and optical spectral observations of O and B-type stars \citep{2015A&A...581A..75K}, Horizontal Branch stars \citep{Jie2023}, hot subdwarfs \citep{Lei2023,Nemeth2021, Luo2020}, and white dwarfs \citep{Wang2022, Vennes2017}. 
It was designed to perform fully automated or supervised spectral analyses of massive data sets.
The procedure starts with an input model and by applying successive approximations along a decreasing global $\chi^2$, iteratively converges on the best solution. 
All models are calculated on the fly, which is the main advantage of the procedure. 
{\sc XTgrid} does not require a precalculated grid and with its tailor-made models -- although at a high computational cost -- it is able to address nonlinearities in multidimensional parameter space. 
After the fitting procedure has converged for a relative change less than 0.5\%, parameter errors are evaluated in one dimension, changing each parameter until the $\chi^2$ variation corresponds to the 60\% confidence limit.
Parameter correlations are evaluated only for effective temperature and surface gravity. 
If the procedure finds a better solution during error calculations, it returns to the descent part using the previous solution as the initial model. 





Instead of using TLUSTY (plane-parallel model, no wind), we used CMFGEN since these stars have non-negligible stellar winds with mass-loss rates in the order of $10^{-8}-10^{-7}$ $M_\odot$ yr\textsuperscript{-1}. 
Using TLUSTY, instead of CMFGEN, would prevent us to address the wind variability and perform homogeneous modeling for BSGs.
In addition, beyond short-term photometric variability, typical for pulsating stars and discussed in Section \ref{frequency_analysis}, gradual spectral variations may occur due to inhomogeneities in the wind density structure.
These together require re-evaluating the surface and wind parameters for each observation.
Therefore, we decided to proceed with CMFGEN models and start out from the results of \citep{2018A&A...614A..91H}. 
We updated {\sc XTgrid} to apply CMFGEN and minimize the wind properties along with the stellar surface parameters. 
Then, we performed a new analysis of the most recent CASLEO spectra to measure the CNO abundances in each of the three stars. 

 The focus of our analysis was on the CNO abundances. Therefore, we kept the abundances of all elements heavier than O at their solar values \citep{Asplund2009} and we adopted He abundances $n{\rm He}/n{\rm H}=$ 0.2, based on the analysis of \citep{searle08}. 
To maintain an approximate consistency with the results of \citep{2018A&A...614A..91H} we kept the stellar radii and turbulent velocity fixed at the values determined in the previous analysis (see Table \ref{starstelpar}) and we applied unclumped wind models. 
Adopting clumping in our models would result in different mass-loss rates.
We neglected macroturbulence in the current analysis because it shows a degeneracy with the projected rotation at low spectral resolutions. 
We note that our goal was to measure the CNO abundances and not to make a comparison with the analysis of \citep{2018A&A...614A..91H}. 
Our method is not suitable for such a comparison, as we used different model atmosphere codes, our observations were taken at different epochs and we used a different fitting procedure. 

\begin{figure}[H]
\begin{adjustwidth}{-\extralength}{-5cm}
\centering
\includegraphics[width=\textwidth]{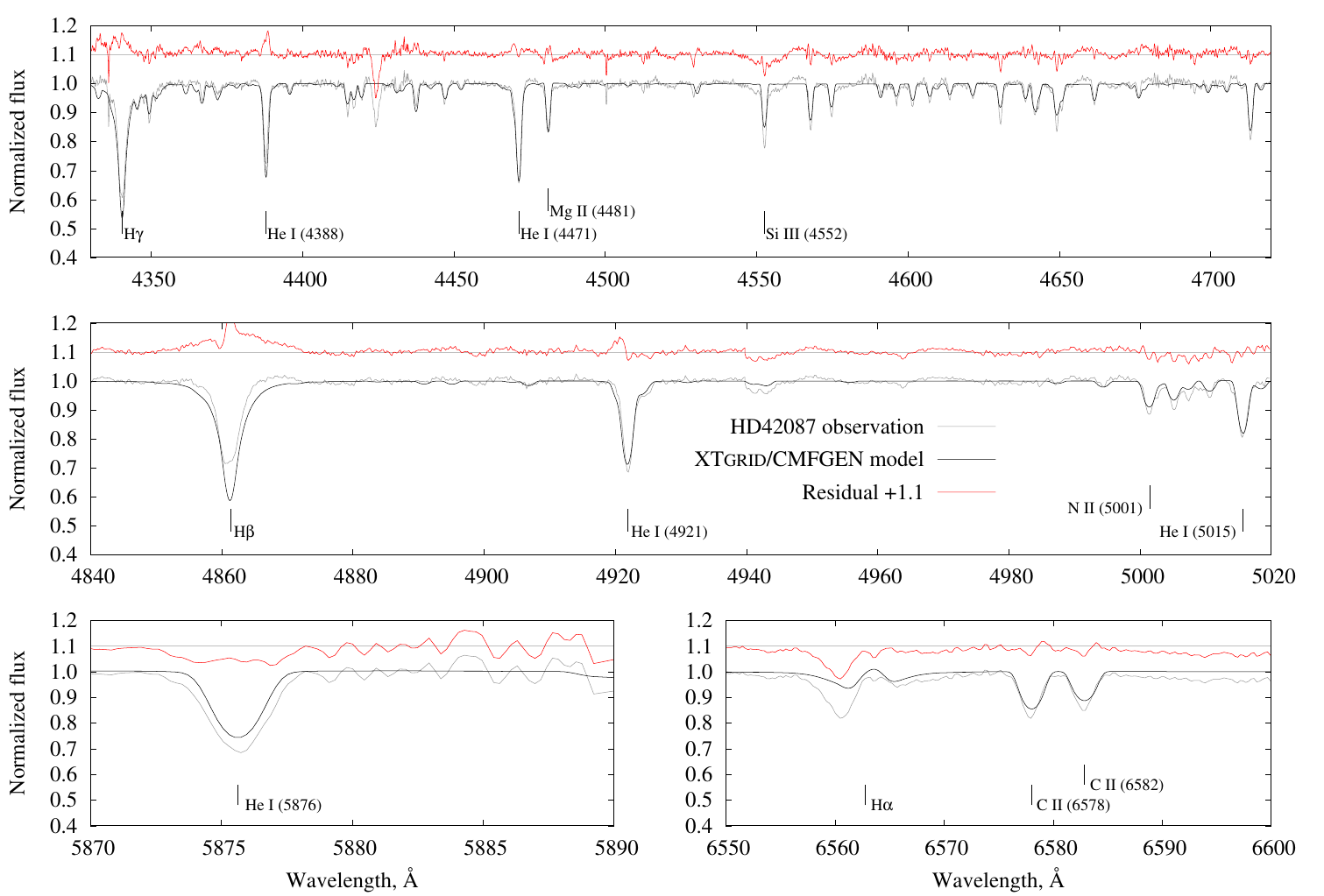}
\end{adjustwidth}
\caption{Best-fit {\sc XTgrid/CMFGEN} model for HD\,42087. 
In each panel, the CASLEO observation is in {\sl grey}, the CMFGEN model in {\sl black}, and the residuals, shifted by +1.1 for clarity, is in {\sl red}. \label{fig_sp_2}}
\end{figure}

\begin{figure}[H]
\begin{adjustwidth}{-\extralength}{-5cm}
\centering
\includegraphics[width=\textwidth]{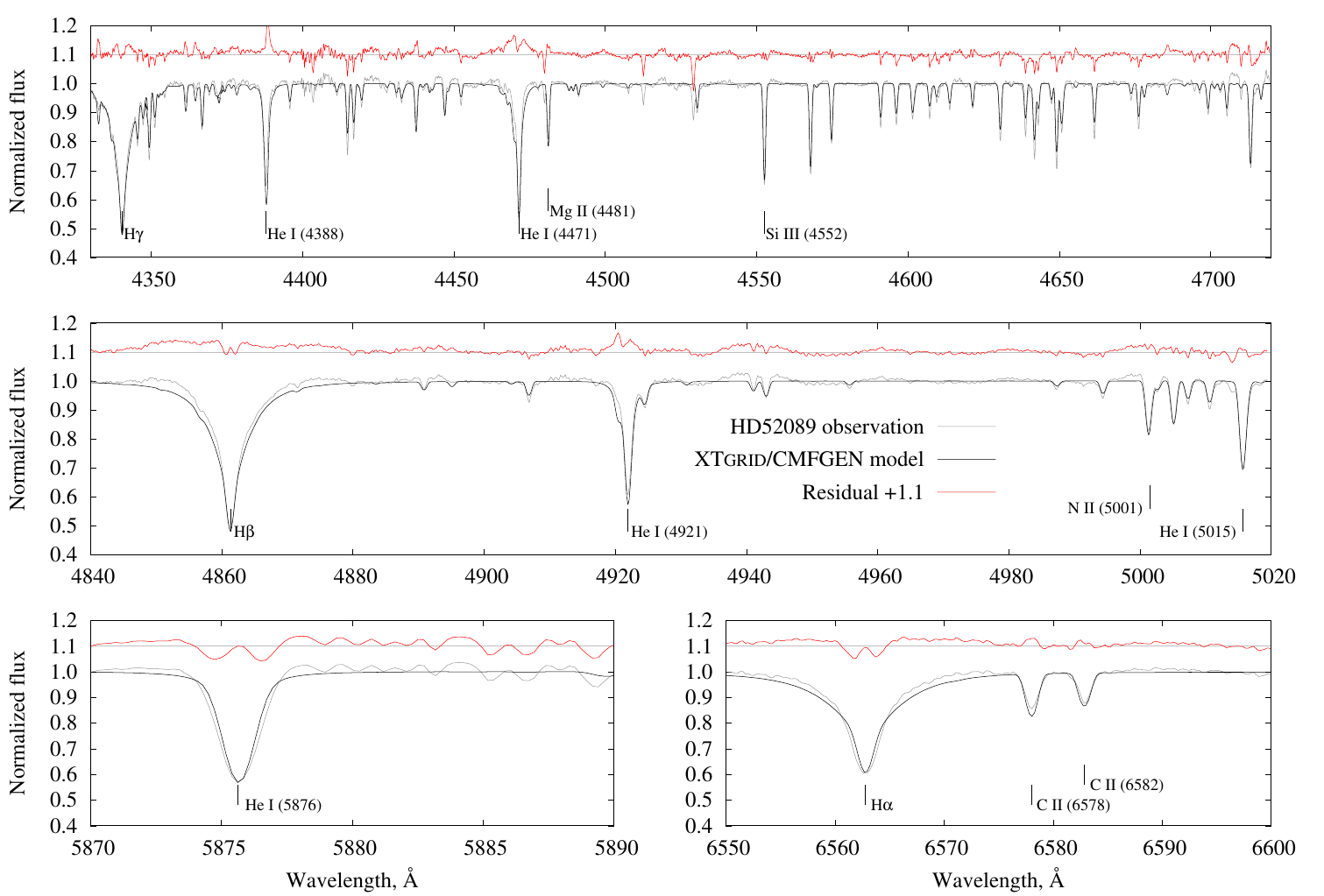}
\end{adjustwidth}
\caption{Best-fit {\sc XTgrid/CMFGEN} model for HD\,52089. 
In each panel, the CASLEO observation is in {\sl grey}, the CMFGEN model in {\sl black}, and the residuals, shifted by +1.1 for clarity, is in {\sl red}. \label{fig_sp_1}}
\end{figure}

\begin{figure}[H]
\begin{adjustwidth}{-\extralength}{-5cm}
\centering
\includegraphics[width=\textwidth]{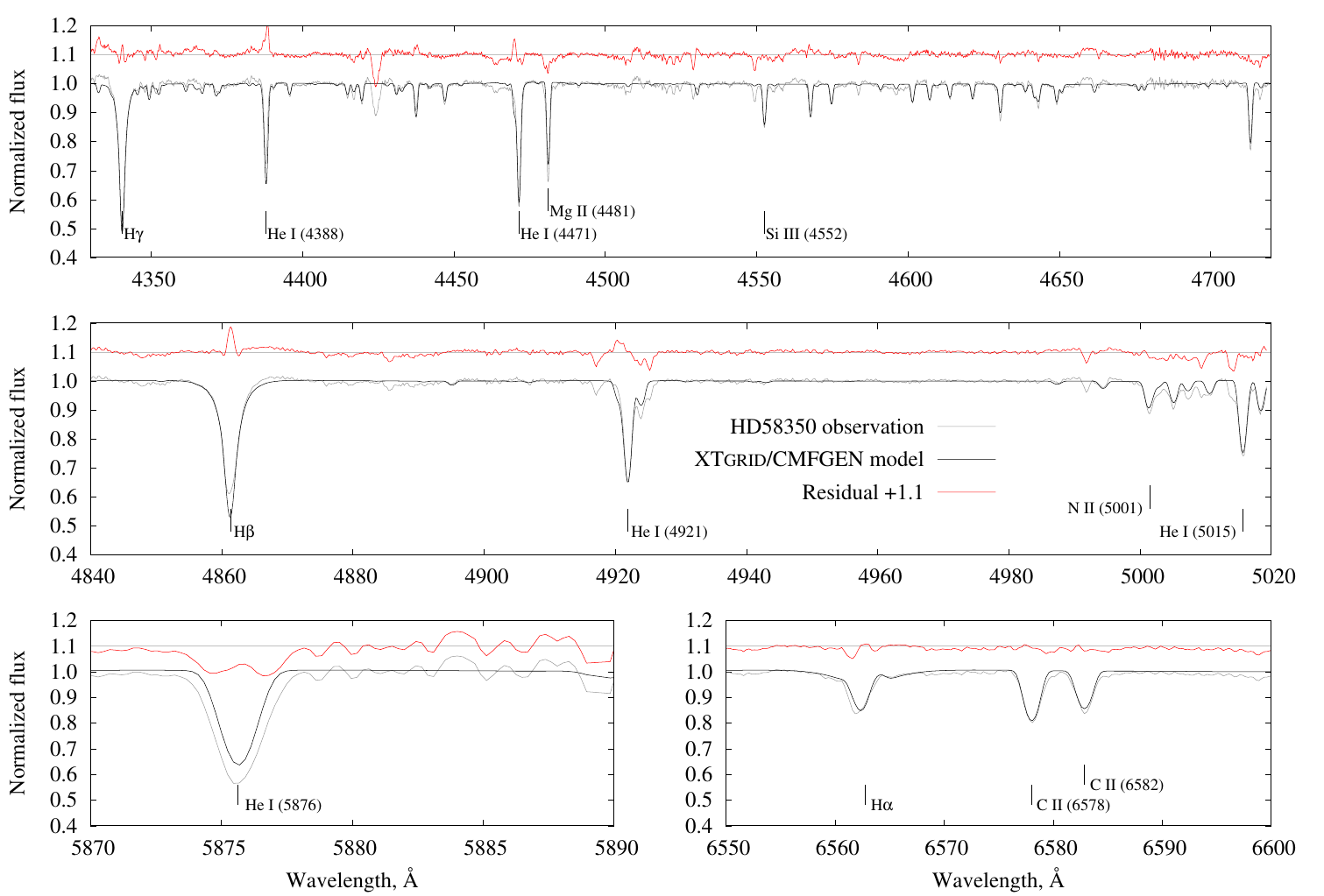}
\end{adjustwidth}
\caption{Best-fit {\sc XTgrid/CMFGEN} model for HD\,58350. 
In each panel, the CASLEO observation is in {\sl grey}, the CMFGEN model in {\sl black}, and the residuals, shifted by +1.1 for clarity, is in {\sl red}. \label{fig_sp_3}}
\end{figure}

Finally, our best-fit CMFGEN models to the CASLEO spectra of HD 42087, HD 52089, and HD 58350, are shown, respectively, in Figs.~\ref{fig_sp_2}, \ref{fig_sp_1} and \ref{fig_sp_3}, and the derived parameters are listed in Table \ref{tab_sp_1}. The luminosity and mass in this table were derived using $L_\star =4 \pi R_\star^2 \sigma T_{\rm eff}^4$ and $g=GM_\star/{R_\star}^2$.

\begin{table}[H] 
\caption{Summary of the spectroscopic results from unclumped CMFGEN models using the {\sc XTgrid} fitting procedure. Abundances are reported on the 12 scale: $\epsilon=\log{N_{\rm x}/N_{\rm H}} + 12$. Fixed parameters are marked with an "x" prefix. Metals not listed here were included at their solar metallicity from \cite{Asplund2009}.\label{tab_sp_1}}
\newcolumntype{C}{>{\centering\arraybackslash}X}
\begin{tabularx}{\textwidth}{lCc|Cc|Cc}
\toprule
\textbf{Parameter} & \multicolumn{2}{c|}{\textbf{HD\,42087}} & \multicolumn{2}{c|}{\textbf{HD\,52089}} & \multicolumn{2}{c}{\textbf{HD\,58350}} \\
\midrule
$T_{\rm eff}$\ (K)& \multicolumn{2}{c|}{18400$^{+1000}_{-\ 200}$} & \multicolumn{2}{c|}{23800$^{+3900}_{-1400}$} & \multicolumn{2}{c}{15800$^{+100}_{-400}$} \\
$\log{g}\ ({\rm cm\ s}^{-2})$ & \multicolumn{2}{c|}{2.34$^{+0.01}_{-0.17}$} & \multicolumn{2}{c|}{3.40$^{+0.01}_{-0.60}$} & \multicolumn{2}{c}{1.95$^{+0.02}_{-0.03}$}\\
$v\sin{i}\ ({\rm km\ s}^{-1})$ & \multicolumn{2}{c|}{73.4 $\pm$ 8.0} & \multicolumn{2}{c|}{38.4 $\pm$ 5.0} & \multicolumn{2}{c}{51.5 $\pm$ 5.0} \\
$v_{\rm turb}\ ({\rm km\ s}^{-1})$ & \multicolumn{2}{c|}{x10} & \multicolumn{2}{c|}{x10} & \multicolumn{2}{c}{x12} \\

$\dot{M}$\ ($M_\odot$ yr\textsuperscript{-1}) & \multicolumn{2}{c|}{(2.3$\pm$1.0)$\times 10^{-7}$} & \multicolumn{2}{c|}{(1.9 $\pm$ 0.2)$\times 10^{-8}$} & \multicolumn{2}{c}{(6.2 $\pm$ 2.0)$\times 10^{-8}$} \\
$v_\infty\ ({\rm km\ s}^{-1})$         & \multicolumn{2}{c|}{x700}   & \multicolumn{2}{c|}{x900} & \multicolumn{2}{c}{x230} \\
$\beta$ & \multicolumn{2}{c|}{x2} & \multicolumn{2}{c|}{x1} & \multicolumn{2}{c}{x3} \\
$L_\star$ ($L_\odot$)  & \multicolumn{2}{c|}{312700$^{+74000}_{-13000}$} & \multicolumn{2}{c|}{35000$^{+29200}_{-\ 7500}$} & \multicolumn{2}{c}{163800$^{+\ 4200}_{-15900}$} \\
$M_\star$ ($M_\odot$)  & \multicolumn{2}{c|}{24.3} & \multicolumn{2}{c|}{11.1} & \multicolumn{2}{c}{9.5} \\
$R_\star$ ($R_\odot$)  & \multicolumn{2}{c|}{x55} & \multicolumn{2}{c|}{x11} & \multicolumn{2}{c}{x54} \\
$\log{L_\star/M_\star}$& \multicolumn{2}{c|}{4.1} & \multicolumn{2}{c|}{3.5} & \multicolumn{2}{c}{4.2} \\
Mean atomic &  \multicolumn{2}{c|}{1.4490} & \multicolumn{2}{c|}{1.5097} & \multicolumn{2}{c}{1.5095} \\
mass (a.m.u.) && && && \\
\midrule
Distance ($pc$)        & \multicolumn{2}{c|}{2470$^{+420}_{-290}$} & \multicolumn{2}{c|}{124$\pm2$} & \multicolumn{2}{c}{608$^{+148}_{-148}$}\\
E($B-V$) ($mag$)         & \multicolumn{2}{c|}{0.4}                  & \multicolumn{2}{c|}{0.005}     & \multicolumn{2}{c}{0.03}             \\
\midrule
{\bf{Element}} & $\epsilon$  &{\bf{mass fr.}}& $\epsilon$ &{\bf{mass fr.}}& $\epsilon$ &{\bf{mass fr.}}\\
\midrule
Hydrogen  & 12              & 5.89$\times 10^{-1}$ &    12           & 5.52$\times 10^{-1}$ &     12          & 5.52$\times 10^{-1}$ \\
Helium    & x11.23$\pm0.10$ & 4.01$\times 10^{-1}$ & x11.30$\pm0.17$ & 4.41$\times 10^{-1}$ & x11.31$\pm0.12$ & 4.41$\times 10^{-1}$ \\
Carbon    & 8.31$\pm0.08$   & 1.37$\times 10^{-3}$ & 8.19$\pm0.15$   & 1.04$\times 10^{-3}$ & 8.07$\pm0.08$   & 7.75$\times 10^{-4}$ \\
Nitrogen  & 8.12$\pm0.06$   & 1.09$\times 10^{-3}$ & 7.97$\pm0.06$   & 7.25$\times 10^{-4}$ & 8.21$\pm0.12$   & 1.25$\times 10^{-3}$ \\
Oxygen    & 8.60$\pm0.08$   & 3.75$\times 10^{-3}$ & 8.30$\pm0.13$   & 1.78$\times 10^{-3}$ & 8.19$\pm0.09$   & 1.38$\times 10^{-3}$ \\
\midrule
                 & [N/C] & [N/O] & [N/C] & [N/O] & [N/C] & [N/O] \\
Abundance ratios & 0.41  & 0.38  & 0.38  & 0.53  & 0.74  & 0.88  \\
\bottomrule
\end{tabularx}
\end{table}




\section{Discussion} 
\label{discussion}

\begin{figure}[H]
\begin{adjustwidth}{-\extralength}{-5cm}
\centering
\includegraphics[width=\textwidth]{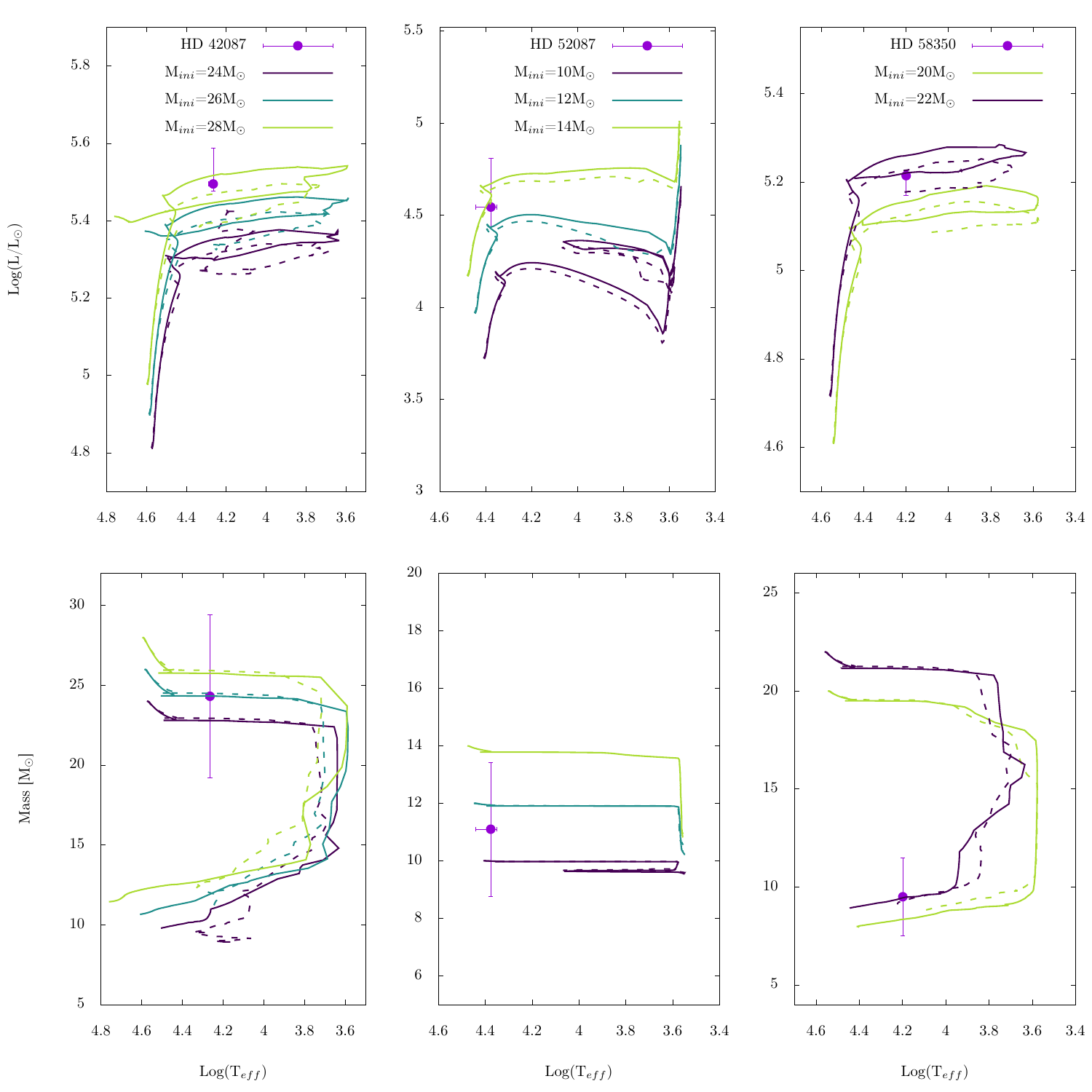}
\end{adjustwidth}
\caption{\textit{Left panels}: HR diagram (top) and mass evolution (bottom) showing the position of HD\,42087 and evolutionary tracks with initial masses of 24, 26, and 28 $M_{\odot}$. \textit{Middle panels}: The same as in the left panels for HD\,52087 and initial masses of 10, 12, and 14 $M_{\odot}$. \textit{Right panels}: the same as before for HD\,58350 and initial masses of 20 and 22  $M_{\odot}$. Solid and dashed lines represent models with $\Omega/\Omega_{crit}=$ 0.568 from \citep{2012A&A...537A.146E}  and the interpolated ones for $\Omega/\Omega_{crit}=$ 0.4, respectively.}\label{HRmass}
\end{figure}

We used the evolutionary sequences from \citet{2012A&A...537A.146E} to explore the evolutionary stage of our star sample. The main physical ingredients of these sequences relevant to our analysis include initial abundances of H, He, and metals set to X = 0.720,Y = 0.266, and Z = 0.014 with chemical initial abundances of C = 2.283$\times 10^{-3}$, N = 6.588$\times 10^{-4}$, O = 5.718$\times 10^{-3}$, in mass fraction. We considered evolutionary tracks with differential rotation at two different rates $\Omega/\Omega_{crit}$ =  0.568, and $\Omega/\Omega_{crit}$ = 0.4, employing for the latter an interpolation of the models with $\Omega/\Omega_{crit}$ = 0.568 and 0, provided in \citet{2012A&A...537A.146E}. The mass loss recipes employed in these sequences are those of \citet{2001A&A...369..574V} for initial masses above 7 $M_{\odot}$. For initial masses above 15 $M_{\odot}$ and $\log(T_{\rm eff})$ > 3.7, \citet{1988A&AS...72..259D} recipe was adopted and the correction factor for the radiative mass-loss rate from \citet{2000A&A...361..159M} was implemented in these rotating models (see, Eq. 10 from  \citet{2012A&A...537A.146E}). Detailed descriptions on the microphysics and mass loss recipes employed in these evolutionary sequencies can be found in \citet{2012A&A...537A.146E} and \citet{2022MNRAS.511.2814Y}. Our selected stars with the derived $T_{\rm eff}$, $\log L_\star$ and mass, along with the evolutionary tracks, are  depicted in Fig. \ref{HRmass}. The errors in this Figure correspond to an error of 10\% in the radii \citep{2018A&A...614A..91H} which in turn result in a 21\% error in the mass and luminosity.

Below, we discuss our results for each star in terms of their location in the HR diagram, their oscillations and the mass-loss rate found and next we compare our results on their surface abundances with previous studies.

\subsection{HD\,42087}

In the left panel of Figure \ref{HRmass} we show the new position of HD\,42087 in an HR diagram along with evolutionary tracks corresponding to initial masses of 24, 26, and 28 $M_{\odot}$. Continuous and dashed lines indicate rotation velocities of $\Omega/\Omega_{crit}$ = 0.568 and $\Omega/\Omega_{crit}$ = 0.4, respectively. In the lower panels, we include a diagram showing the total mass evolution. The derived values for $T_{\rm eff}$ and $\log (L_\star/L_{\odot})$ suggest an initial mass of $\sim$26  $M_{\odot}$ and our value obtained for the current mass indicate this star is at the pre-RSG stage, which is in agreement with the mass evolution diagram for stars with the mentioned initial mass.

\begin{figure}[H]
\begin{adjustwidth}{-\extralength}{-5cm}
\centering
\includegraphics[width=\textwidth]{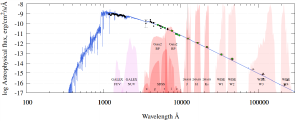}
\end{adjustwidth}
\caption{SED of HD\,42087. 
All data points were taken from the VizieR Photometry Viewer service.
The photometric data were de-reddened using $E(B-V)$ = 0.4 mag. 
The green points were used to match the slope of the passband convolved CMFGEN fluxes to the observations and the model was normalized to the observed SED in the 2MASS/J band.  The binned IUE spectrum is included with black dots.
}\label{HD42_sed} 
\end{figure}

Our analysis for this object resulted in a higher $T_{\rm eff}$ and lower $\log{g}$ than those from \citet{searle08} and \citet{2018A&A...614A..91H}. We emphasize that \citet{2018A&A...614A..91H} did not have the Si lines covered by their spectra and in principle, our new values would be more reliable. Additionally, with our values, HD\,42087 lies in the linear relation ($\log T_{\rm eff} - \log{g}$) found in \citet{searle08} for Galactic BSGs (their Fig. 6). In Fig.\,\ref{HD42_sed}, we show the theoretical SED using the D = $2470^{+420}_{-289}$ pc Gaia EDR3 distance and $E(B-V)$ = 0.4 mag extinction adopted from the {\sc Stilism} maps \citep{Capitanio2017}, along with the binned IUE spectrum in black. Our procedure uses the fit formula from \citet{1989ApJ...345..245C} based on the extinction data from \citet{1986ApJ...307..286F,1988ApJ...328..734F} with R$_v$=3.1. The theoretical SED fits the photometric measurements, including the IUE spectrum, and the resulting SED and CMFGEN model masses agree within error bars , being 22.5 M$_{\odot}$ the mass derived from the SED. However, we were unable to match the H$\alpha$ and H$\beta$ profiles with our homogeneous wind models. 
Considering the reported line variability and asymmetries by \citet{morel04}, which is obvious also from Fig. \ref{fig_HD42_alpha}, we conclude that our steady, smooth wind model is inadequate for the 2020 CASLEO spectrum of HD\,42087.

Nevertheless, we were able to derive a mass-loss rate of $\dot{M}$ = $2.3\times 10^{-7}$ $M_\odot$ yr\textsuperscript{-1} for this star with our procedure. This value is close to the value found by \citet{2018A&A...614A..91H} of $\dot{M}$ = $5.7\times 10^{-7}$ $M_\odot$ yr\textsuperscript{-1}, being lower by a factor of $\sim$2.6. From looking at their observed H$\alpha$ line profile (taken in 2006, Fig.\ref{fig_HD42_alpha}), we see a stronger emission component in H$\alpha$ compared to our data (2020). This is in line with a higher mass-loss rate (in 2006) as found by these authors.

The frequency analysis performed for HD\,42087 shows three frequencies, $f_1$, $f_2$ and $f_3$ which appear randomly over the three sectors and two more detectable in the long time span of the three consecutive sectors ($f_4$ and $f_5$). The short lifetime for these variations prevents us from associating them with stellar pulsations but rather with stochastic variations at the stellar surface \citep{2021A&A...648A..79K}. Additionally, the analysis over the combined sectors allowed us to find one stable frequency, $f_r$, possibly associated with stellar pulsations. The nature of this mode is not certain. Its frequency lies in the usual range ([10, 100] d) of strange modes, known to facilitate the mass loss in massive stars \citep{2010A&A...513L..11A}. These modes can be radial and non-radial \citep{2011MNRAS.412.1814S} and they appear trapped in the strongly inflated envelopes of highly non-adiabatic stars, usually with $\log L_{\star}/M_{\star} > 4$. The low amplitude of $f_r$ and the new value found for $\log L_{\star}/M_{\star} = 4.1$, indicate that this can be a non-radial strange mode. If true, it would explain the high variability in the wind lines (Fig. \ref{fig_HD42_alpha}) and why the spectrum can not be modelled with a smooth wind approach. Nevertheless, the only way to determine if this mode facilitates the mass loss in HD\,42087 is to perform a nonlinear stability analysis and to check whether the mode velocity on the surface can exceed the stellar escape velocity. We did not detect any frequency corresponding to the  $\sim$25 d period observed for the variability in the H${\alpha}$ line. We stress here that the detection of any periodicity from the H$\alpha$ profile in the light curves is unlikely to be observed. Any perturbation on the stellar surface produced in the large wind volume where the H$\alpha$ line is formed may lead to new perturbations (for example due to densities inhomogeneities in the wind), difficulting its detection. Additionally, the H$\alpha$ profile, in most cases is composed of absorption and emission components and the product of both components in the integrated flux also difficult its detection. However, as shown in  \citet{2021A&A...648A..79K}, wind variations can cause stochastic light variations if the base perturbations are sufficiently large. The lack of multiple independent frequencies identified as stellar pulsations modes in this object would indicate that this star belongs to the pre-RSG stage, considering that massive post-RSG have in general more excited modes than pre-RSG, as shown in \citet{2013MNRAS.433.1246S}. This is in agreement with our values for $T_{\rm eff}$, $\log g$, $M_{\star}$ (Fig. \ref{HRmass}).

\subsection{HD\,52089}


The new position of HD\,52089 in the HR diagram along with evolutionary tracks for initial masses  10, 12 and 14 $M_{\odot}$ and $\Omega/\Omega_{crit}$ = 0.568 and 0.4 (continuous and dashed lines, respectively) are depicted in Fig. \ref{HRmass}. The HR diagram suggests initial masses between 12 and 14 M$_{\odot}$ at the TAMS in agreement with our derived value for the current mass ($\sim 11M_{\odot}$) in this stage considering the adopted errors in the mass. However, we can not dismiss a merger scenario for this object that would explain the measured luminosity, which is slightly high considering an 11 $M_{\odot}$ object at the TAMS. Such an event can lead to rejuvenation and an overluminosity of the merger remnant due to the energy injection from the secondary during the merging of the two components. The merger product might hence appear as if it would have a significantly higher mass. The best known such case is the B[e] supergiant star R4 in the Small Magellanic Cloud \citep{1996A&A...309..505Z, 2020ApJ...901...44W}. If true, then HD 52089 would be a highly important  object to study merger remnants, and it would be interesting to search for possible remnants of ejecta from the past merger event.



\begin{figure}[H]
\begin{adjustwidth}{-\extralength}{-5cm}
\centering
\includegraphics[width=\textwidth]{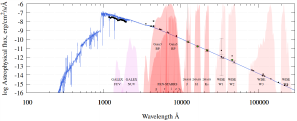}
\end{adjustwidth}
\caption{SED of HD\,52089. 
Same as Fig.\,\ref{HD42_sed}, but de-reddening was done using $E(B-V)$= 0.005 mag. 
\label{HD52_sed} }
\end{figure}

Additionally, this star showed an inconsistency in the $T_{\rm eff}-\log{g}$ distribution of \cite{2018A&A...614A..91H}, having a higher $\log{g}$ for its temperature than other stars in their sample.  Our new fit to the 2015 CASLEO spectrum confirmed the earlier results and shows a discrepancy among the stellar mass, luminosity, and surface gravity, given its high effective temperature. 
With its close distance of D = $124\pm2$ pc and moderate interstellar extinction of $E(B-V)$ = $0.005$ mag, adopted from \citep{2007A&A...474..653V} and {\sc Stilism}, respectively, we find that its spectroscopic mass is in good agreement with the mass derived from the SED (9.8M$_{\odot}$) depicted in Fig. \ref{HD52_sed}.  We also notice the model from the optical fit matches the slope of the binned IUE spectrum, but there is an offset possibly due to the low metalicity found for this object.

The frequency analysis for HD\,52089 over the 4 TESS sectors revealed one triplet centred in $f_2$ in Sector 33, corresponding to a rotational period of 5 d ($f_{rot}$ = 0.2 d$^{-1}$). The short baseline observations provided by TESS single sectors translate into high Raylight frequency separation, preventing making precise mode identification for short-term variabilities. In the case of $f_3$, $f_2$ and $f_4$ being members of a triplet, $f_2$ would be a a $\ell$ = 1 nonradial mode \cite{2013MNRAS.433.1246S}. In this case, the star would have a surface rotation velocity of $\sim$110 km s$^{-1}$, considering the radius and rotation period derived for this object. Moreover, with the value of v $\sin i$ obtained from our analysis, we can derive the inclination of the star, which would be $\sim$20 degree. Besides the triplet, we found $f_2$ in Sectors 7 and 33, and signatures of this mode in Sectors 6 and 34 with very low amplitude (not included in Table \ref{tab10}) along with its harmonic supporting the hypothesis of $f_2$ being a non-radial mode due to its apparently extended lifetime. As in the case of HD\,42087, we found low-frequency signals ($f_1$, $f_5$ and $f_6$) randomly excited over the observed sectors, which are probably related to convective variabilities stochastically excited at the stellar surface. These observations confirm the results of \cite{Brussens2020}.



 Our determination of $\dot{M}$ for HD\,52089, $\dot{M}$ = $1.9\times 10^{-8}$ $M_\odot$ yr\textsuperscript{-1}, agrees very well with the one reported by~\citet{2018A&A...614A..91H}: $\dot{M}$ = $2.0\times 10^{-8}$ $M_\odot$. From comparing our observed H$\alpha$ line profile of this star, we do not find any significant morphological difference between the observed spectrum shown in ~\citet{2018A&A...614A..91H} and our data: a pure absorption H$\alpha$ line profile with flux at the core of the line of $\sim$0.6 (normalized flux) (see Fig. \ref{fig_HD52_alpha}).\par

\subsection{HD\,58350}

The right panel of Fig. \ref{HRmass} shows the position of HD\,58350 in the HR diagram and the evolution of the total mass along with evolutionary tracks for 20 and 22 $M_{\odot}$ with different rotational velocities. We found an excellent agreement between our derived values for $T_{\rm eff}$, $\log (L_\star/L_{\odot})$ and its current mass for a star evolution model with initial mass $\sim$22 M$_{\odot}$ at the post- RSG stage, after losing a considerable amount of mass during its evolution. We notice as well, that our derived values for the $T_{\rm eff}$ and $\log g$ are in good agreement with the linear relation found for galactic BSGs in \citet{searle08}.

\begin{figure}[H]
\begin{adjustwidth}{-\extralength}{-5cm}
\centering
\includegraphics[width=\textwidth]{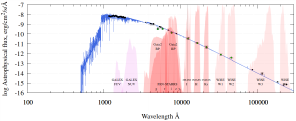}
\end{adjustwidth}
\caption{SED of HD\,58350. 
Same as Fig.\,\ref{HD42_sed}, but de-reddening was done using $E(B-V)$ = 0.18 mag (see main text).
\label{HD58_sed} }
\end{figure}

The theoretical SED fits the optical photometric measurements for HD\,58350 (Fig. \ref{HD58_sed}). We notice a slightly different slope for the IR photometric data, possibly pointing towards a time-variable wind in data taken at different dates.  Considering its distance $D=$ 608 $\pm$ 148 pc \citep{2018A&A...614A..91H}, we derived an extinction $E(B-V)$ = 0.18 mag. We also found a discrepancy between the stellar masses obtained from the SED modeling ($\sim$5.5 $M_{\odot}$) and the best-fitting CMFGEN model ($\sim$9.5 $M_{\odot}$) of unclear origin. We have found different values for the effective temperature (and extinction) in the literature (see Sec. \ref{Peter}) for this object, ranging from 13500 K to 16000 K, possibly due to a combination between the use of different methodologies to derive it and due to stellar oscillations. This hampers a reliable comparison between our SED model and the photometric data, possibly leading to this mass discrepancy. Spectroscopic time-series observations analyzed homogeneously can help to place reliable constraints on the effective temperature.

 For HD 58350, our derived $\dot{M}$ of $6.2\times 10^{-8}$ $M_\odot$ yr\textsuperscript{-1} is lower than the one from~\citet{2018A&A...614A..91H} by a factor of $\sim$2.4. This can be understood since the observed H$\alpha$ line profile reported in ~\citet{2018A&A...614A..91H} (taken in 2006 and 2013) shows a more intense emission component in comparison with our data (very weak emission in H$\alpha$).

The frequency content for this star lies below 0.4 d $^{-1}$. The only sector provided by TESS at the current time allowed us to find two independent frequencies. We were not able to find the frequency discovered in \citet{2007A&A...463.1093L}, with $f$ = 0.1507 d$^{-1}$, indicating that this mode is not an opacity-driven mode. From the analysis of one single TESS sector, we can not suggest this star is a non-radial oscillator. However, the marked H$\alpha$ line profile variation (see Fig. \ref{fig_HD58_alpha}) suggests these frequencies are connected to line-driven wind instabilities \citep{2021A&A...648A..79K}. The analysis in \citep{2013MNRAS.433.1246S} demonstrated that a BSG in a post-RSG state should undergo multiple pulsations. It is therefore unfortunate that HD\,58350, which is the best candidate in our sample for a post-RSG star, was observed only in one TESS sector. Only with multiple, and in particular consecutive TESS observations, would it be possible to properly analyze the frequency spectrum of this object and to confirm the theoretical predictions of \citet{2013MNRAS.433.1246S}.

\subsection{Surface abundances}

\begin{figure}[H]
\begin{adjustwidth}{-\extralength}{-5cm}
\centering
\includegraphics[width=1\textwidth]{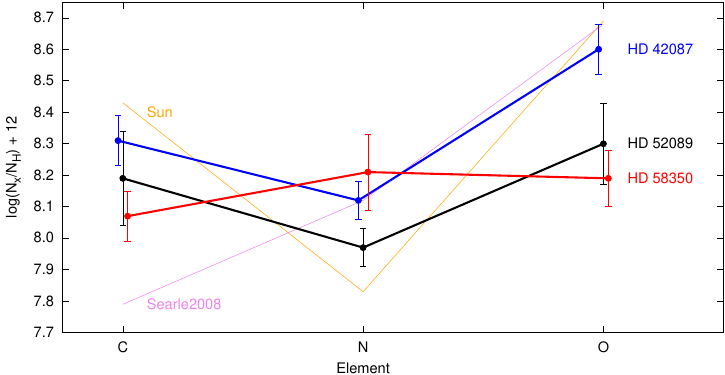}
\end{adjustwidth}
\caption{The measured CNO abundance patterns in all three stars compared to the solar pattern from \citet{Asplund2009}.
In all three stars, C and O are depleted and N is overabundant compared to the solar mixture.
The mean CNO abundances found by \citet{searle08} for Galactic BSGs are also shown for reference.
\label{fig_abn}}
\end{figure}

With the aim of framing these discrepancies with our derived values for the surface abundances, we compared our results with those from different authors and Geneva evolutionary tracks. Fig.~\ref{fig_abn} shows our measured CNO abundances for the selected stars along with the solar abundances and the average CNO abundances derived in \citep{searle08} for their Galactic BSG sample.

In general, for all three stars, we found well-constrained CNO abundances, all showing a slight depletion of C and O, and a mild overabundance of N when compared to the solar mixture. This pattern agrees with the previously reported CNO abundance profiles in \citep{searle08}; however, at the same time we found higher C and slightly lower O abundances.
 We notice that some systematic differences are expected from the analysis itself. The global spectral modeling applied in {\sc XTgrid} is fundamentally different from the methodology of \citet{searle08} who used different diagnostic lines for $T_{\rm eff}$, $\log{g}$ and abundance determinations. 
\citet{searle08} noted that their CMFGEN models, with derived C abundances from the \ion{C}{ii} 4267 \AA\ line, overestimated the \ion{C}{ii} 6578 and 6582 \AA\ lines. 
In contrast, our C abundance analysis was based on the strongest C features in the CASLEO observations (\ion{C}{ii} 6578 and 6582 \AA\ lines), but the global analysis is also sensitive to variations in all other spectral lines due to a change in the C abundance. 
This difference in diagnostics is applied consistently for all surface and wind parameters in {\sc XTgrid}.  
Furthermore, \cite{Nieva2006} showed that LTE analyses could result in discrepant abundances based on different carbon lines, and the \ion{C}{ii} 4267 multiplet tends to underestimate the carbon abundance. 
Further possible sources of a discrepancy may be the differences in atomic data used in the analyses and the different fit procedures. Additionally, we note that our sample of three stars is too small to match \citep{searle08} population averages and the individual objects in our selection may show large deviations.
To uncover such systematics one needs to process larger, homogeneously modeled datasets and multi-epoch observations, which are beyond the scope and limitations of our current work.  

At the same time, we noticed that the total amount of metals we derived (see Table \ref{tab_sp_1}) correspond to subsolar metallicities, being approximately $70\%\ Z_{\odot}$, $60\%\ Z_{\odot}$, and $50\%\ Z_{\odot}$ for HD\,42087, HD\,52089, and HD\,58350 respectively, possibly due to the adopted fixed value for $n{\rm He}/n{\rm H} =0.2$. We have also adopted fixed solar abundances for all metals beyond oxygen and  did not include elements that do not contribute to the opacity, such as Ar, K, Ti, V, Mn, and Co. All these features contribute to reducing the metal mass fractions.

\begin{figure}[H]
\begin{adjustwidth}{-\extralength}{-5cm}
\centering
\includegraphics[width=\textwidth]{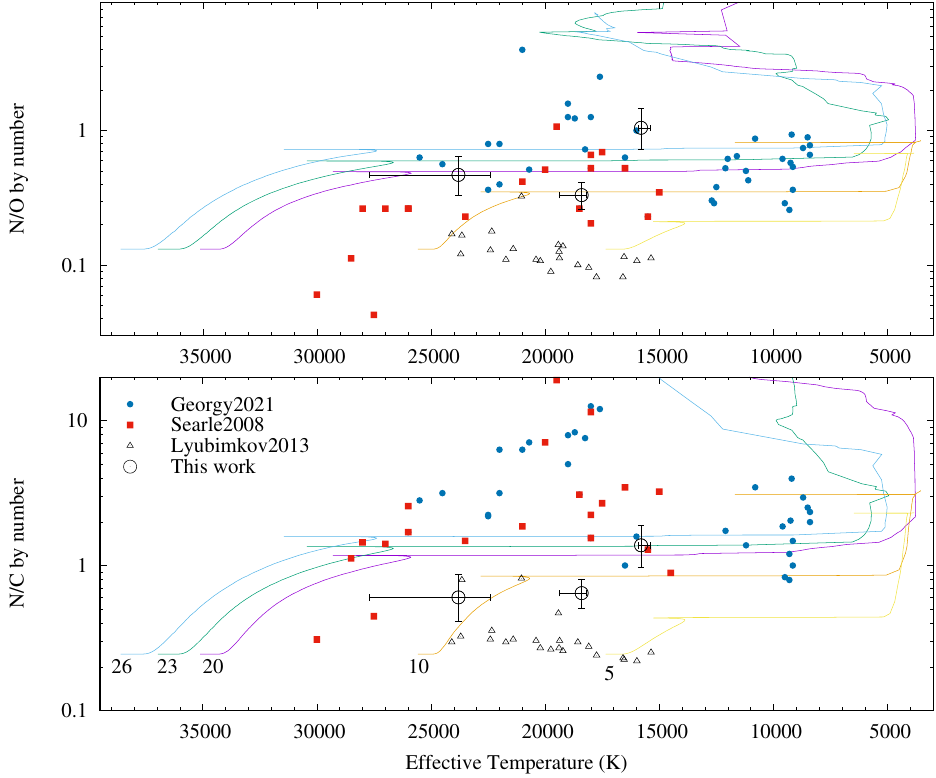}
\end{adjustwidth}
\caption{N/C ($bottom$) and N/O ($top$) abundance ratios by number from \citet{searle08} and \citet{Georgy2021} compared to our measurements and Geneva solar metallicity (Z = 0.014), 
$\Omega/\Omega_{crit}=$ 0.4 interpolated evolutionary tracks for 5, 10, 20, 23 and 26 $M_\odot$. 
For reference, we also show B-type main sequence stars from \citet{Lyubimkov2013}. \label{fig_abnr}}
\end{figure}

In Fig.\,\ref{fig_abnr} we compare the N/C and N/O abundance ratios from different star samples with Geneva stellar evolution models for different masses. We included the sample of A and B supergiants studied in \citep{searle08} and \citep{Georgy2021} and the B-type main sequence (MS) stars analyzed in \citep{Lyubimkov2013} as a reference, along with our three stars.  Three well-defined groups can be identified in the panels of Fig.\,\ref{fig_abnr}. B-type MS stars between 5 and 10 $M_\odot$ form a clear sequence in temperature and show the lowest abundance ratios, which marks that these stars have low N abundances relative to C and O. Below \textasciitilde13000 K, cooler A-type supergiant stars show up as a compact group, while the hot side of the observed BSG abundance ratios shows a much larger scatter. The larger scatter can be interpreted in several ways. \citet{2021A&A...647A..28K} have shown that winds are driven mostly by C, Si, and S for hot BSGs and iron for cooler BSGs implying a decreasing mass-loss rate for temperatures lower than 15000 K. A lower mass-loss rate might operate in favor of a better determination of N/C and N/C ratios. The observed compact group could be as well a signature of the fast post-RSG evolution through the cool BSG stage. Additionally, atmospheres of cool supergiants usually exhibit more lines, helping to obtain precise values in their abundances. However, it is clear that none of the measurements reaches the predicted high N/C and N/O ratios for the post-RSG domain.


Based on the abundance ratios alone, we find that all our stars are more consistent with a pre-RSG stage of evolution, we do not see the predicted very high N/C and N/O ratios. 
This is in contrast with the spectroscopic mass of HD\,58350, which suggests it is in a post-RSG stage. 
\citet{2013MNRAS.433.1246S} showed the N/C and N/O ratios are increased mainly by the mass loss. However, they found N/C and N/O abundances consistent with models at the pre-RSG stage, in contradiction with the position in the HR diagram for Deneb and Rigel, arriving at the same conclusion.

There is a discrepancy in the abundance ratios between evolutionary model predictions and spectroscopic measurements.  
The observed ratios remain significantly lower than predictions. Much larger samples, multifaceted efforts, and homogeneous modeling will be necessary to address this issue statistically.

\begin{figure}[H]
\begin{adjustwidth}{-\extralength}{-5cm}
\centering
\includegraphics[width=0.8\textwidth]{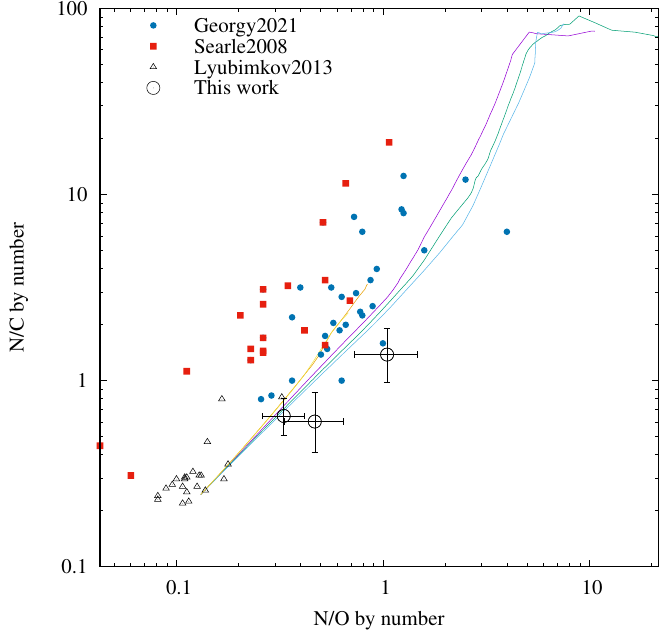}
\end{adjustwidth}
\caption{N/C and N/O abundance ratio correlations based on the same datasets as in Fig.\,\ref{fig_abnr}. \label{fig_abnrc}}
\end{figure}

Fig.\,\ref{fig_abnrc} shows the correlations between the N/O and N/C ratios, which is analogous to the distribution  \citep{Martins2015} found for O-type stars.

The offset between theoretical predictions and the measurements implies that some systematics may exist in the C and O abundances. 
A systematically underestimated C or overestimated O abundance can produce the observed offset. 
It is unlikely that all the methodologies used to analyze the surface abundances in these stars are biased in the same way, therefore, the origin of the offset remains unclear. 
It may be related to stellar variability as well as to atomic data or shortcomings in the surface abundance predictions.
\citet{Martins2015} demonstrated an anticorrelation between the N/C ratio and $\log{g}$, the lower the gravity the larger the N/C ratio. 
Meanwhile, \citet{2013MNRAS.433.1246S} concluded that recent developments in modeling of RSG \citep{2013ApJ...767....3D} make these stars more compact for a given luminosity. 
The combination of the two trends acts towards decreasing the offset with respect to the Geneva models in Fig.\,\ref{fig_abnrc}.

\section{Conclusions}
\label{conclusions}

This paper reports our first step towards a comprehensive study of BSG stars, taking into account their photometric and spectroscopic variabilities.

The evolutionary tracks predict HD\,42087 to be a pre-RSG star for our derived values of $\log (L_\star/L_{\odot})$, $T_{\rm eff}$, and $M_\star$, in agreement with our values for the abundance ratios.  HD\,52089 is most likely an 11 $M_{\odot}$ star at the TAMS. However, we noticed that the derived luminosity is slightly high for an 11 $M_{\odot}$ star, and a binary merger scenario is plausible. Finally, for HD\,58350, the evolutionary tracks for our derived values indicate this star is at the post-RSG. However, the derived surface abundance ratios are compatible with those at the pre-RSG stage, finding the same discrepancies mentioned in \citep{2013MNRAS.433.1246S} for Deneb and Rigel.
The rather short sectoral observing windows of TESS are insufficient to cover the low frequencies usually present in these stars. However, we were able to detect a frequency splitting allowing us to infer a rotational period of $\sim$5 d for HD\,52089 and to find a low frequency, $f_r$ = 0.093 $d^{-1}$, related possibly to strange mode instabilities in HD\,42089 in agreement with its new derived value for the $\log (L_\star/L_{\odot})$.

We found lower values for mass-loss rates of HD 42097 and HD 58350 compared with those derived by \citet{2018A&A...614A..91H}, in agreement with the detected variability of the H$\alpha$ emission component, related to changes in the wind mass-loss rate in these stars at these different epochs. The finding of low-frequencies stochastic oscillations in these objects is possibly connected with such variations.
 
%
Despite the numerous and valuable efforts to study BSG stars, there are yet many issues to address and we highlight here the most important ones:
\begin{itemize}
 
\item Large sample of BSGs needs to be studied with homogenously modeled datasets of multi-epoch observations, with the aim of uncovering systematic deviations of surface abundance ratios from evolutionary models. Additionally, multi-epoch observations will allow us to place constraints to the current uncertainties observed in the radii of these objects and to identify changes in radii caused by radial pulsations.

\item To frame the current studies for the needed M$_{core}$/M$_{r}$ ratio for massive stars to evolve back towards the blue region of the HR diagram, such as the effect of stellar rotation, convective boundaries criteria, mixing length theories, overshooting adopted for the stellar interior and evolutionary models, in terms of the CNO abundances. This will allow us to untackle the observed indetermination in the evolutionary stage of these objects with precise values for their CNO surface abundances and, in turn, will help to set the needed constraints to the current poorly established theoretical mass-loss recipes in the diverse evolutionary states and mass ranges.

\item In addition, we emphasize that stellar pulsations play a key role in the analysis of BSGs, not only as a test to infer their evolutionary stage as proposed in \citep{2013MNRAS.433.1246S}, but also as a mechanism that facilitates the mass loss in massive stars as suggested in \citep{2015A&A...581A..75K} and theoretically confirmed in \citep{2016MNRAS.457.4330Y}, affecting, therefore, their surface abundances. The systematic differences noticed when comparing evolutionary tracks with surface abundance measurements for BSGs (Fig. \ref{fig_abnrc}) should be discussed, considering the effect of stellar pulsations over their evolution. Furthermore, short-term mass-loss variabilities should be contemplated in detailed evolutionary sequences as they can act as an additional source for the discrepancies found with evolutionary models.
   
\end{itemize}

From a spectral analysis software development point, we have added CMFGEN modeling capabilities to the automatic spectral analysis procedure {\sc XTgrid}\footnote{\url{https://xtgrid.astroserver.org/}}. The first results shown here demonstrate its feasibility for processing spectra of massive stars and deriving homogeneous parameters from diverse data. The next steps will include improving the accuracy of spectral inference to reduce the observed discrepancies in mass and surface abundances. Additionally, we will work on better optimizing the calculations by recycling previously calculated models. Future applicability for large datasets relies on the development of efficient methods to search the parameter space either by utilizing large grids \cite{Zsargo2020} and/or machine learning techniques. 

We would like to mention as well, that we found diffuse interstellar bands (DIB) in the spectra of HD\,42087 and HD\,58350. 
The 4428 and 6613 \AA\ DIB bands are clearly present in both stars. Neither of the two bands is visible in the spectrum of the relatively nearby HD\,52089. 
Although the two stars with DIBs are at larger distances, and they have very different extinction values, the DIBs show very similar strengths. 
The lack of DIBs in HD\,52089 is likely due to its short distance and the very low interstellar extinction in its direction. 
In addition, HD\,52089 is not only the hottest star in our sample, but also the strongest ultraviolet source in the night sky \citep{Gregorio2002}, which may be able to photodissociate DIB carriers. 

\authorcontributions{``Conceptualization, J.P.S.A, P.N. and M.K; methodology, J.P.S.A. and P.N., E.S.G.A; software, P.N.; validation, P.N. and E.S.G.A; formal analysis, P.N.; investigation, J.P.S.A, P.N., E.S.G.A, M.K. and M.A.R.D.; resources, M.A.R.D and M.H.; writing---original draft preparation, J.P.S.A., P.N., M.A.R.D., M.K. and E.S.G.A.; writing---review and editing, J.P.S.A., P.N., E.S.G.A and M.K.; visualization, J.P.S.A, P.N., M.K, and M.A.R.D; supervision, J.P.S.A.; project administration, J.P.S.A and P.N. All authors have read and agreed to the published version of the manuscript."}

\funding{
This research received funding from the European Union's Framework Programme for Research and Innovation Horizon 2020 (2014-2020) under the Marie Sk\l{}odowska-Curie Grant Agreement No. 823734  (POEMS project). 
The Astronomical Institute in Ond\v{r}ejov is supported by the project RVO:67985815. J.P.S.A and M.K. acknowledge financial support from the Czech Science foundation (GACR 20-00150S).
E.~S.~G.~de Almeida has been financially supported by ANID Fondecyt postdoctoral grant folio \textnumero~3220776.
P.N. acknowledges support from the Grant Agency of
the Czech Republic (GA\v{C}R 22-34467S). 
M.A. Ruiz Diaz acknowledges support from CONICET (PIP 1337)
}

\acknowledgments{We are grateful to the referees for their thoughtful reports and useful suggestions that helped us improve the manuscript. We thank Dr. Cidale for the inspiring and valuable discussions. 
Based on observations taken with the J. Sahade Telescope at Complejo Astronómico El Leoncito (CASLEO), operated under an agreement between the Consejo Nacional de Investigaciones Científicas y Técnicas de la República Argentina, the Secretaría de Ciencia y Tecnología de la Nación, and the National Universities of La Plata, Córdoba, and San Juan.
This research has used the services of \mbox{\url{www.Astroserver.org}} under reference M1DE05. 
}






\appendixtitles{no} 
\appendixstart
\appendix

\reftitle{References}


\bibliography{References}

\end{document}